\documentclass{aa}  

\pdfoutput=1
\usepackage[varg]{txfonts}
\usepackage{graphicx}
\usepackage{natbib}
\usepackage{array,multirow,makecell} 
\usepackage{amssymb}
\usepackage{amsmath}
\usepackage{pifont}
\usepackage{subcaption,caption}
\newcommand{\cmark}{\ding{51}}%
\newcommand{\xmark}{\ding{55}}%
\usepackage{color}
\usepackage{indentfirst}
\usepackage[breaklinks, colorlinks, citecolor=blue]{hyperref}
\bibpunct{(}{)}{;}{a}{}{,} 
\def\mean#1{\left< #1 \right>}

\begin{document} 

\title{Observational constraints on key-parameters of cosmic reionisation history}

\author{A. Gorce \inst{1,}\inst{2}
\and M. Douspis \inst{1}
\and N. Aghanim \inst{1}
\and M. Langer \inst{1}
                }

\institute{Institut d'Astrophysique Spatiale, Université Paris-Sud, CNRS, UMR8617, 
                        91405 Orsay, France\\
                        \email{adelie.gorce@ias.u-psud.fr}
                \and
                        Department of Physics,
                        Blackett Laboratory, 
                        Imperial College,                       
                        London SW7 2AZ, U.K.\\
                }

   \date{Received *********; accepted *******}

\abstract {We discuss constraints on cosmic reionisation and their implications on a cosmic star formation rate (SFR) density $\rho_\mathrm{SFR}$ model; we study the influence of key-parameters such as the clumping factor of ionised hydrogen in the intergalactic medium (IGM) $C_\ion{H}{II}$ and the fraction of ionising photons escaping star-forming galaxies to reionise the IGM $f_\mathrm{esc}$. Our analysis has used SFR history data from luminosity functions, assuming that star-forming galaxies were sufficient to lead the reionisation process at high redshift. We have added two other sets of constraints: measurements of the IGM ionised fraction and the most recent result from Planck Satellite about the integrated Thomson optical depth of the cosmic microwave background (CMB) $\tau_\mathrm{Planck}$. Our analysis shows that a reionisation beginning as early as $z\geq14$ and persisting until $z\sim6$ is a likely scenario.

We also considered various possibilities for the evolution of $f_\mathrm{esc}$ and $C_\ion{H}{II}$ with redshift, and confront them with observational data cited above. We conclude that, if the model of a constant clumping factor is chosen, the fiducial value of three is consistent with observations; even if a redshift-dependent model is considered, the resulting optical depth is strongly correlated with $C_\ion{H}{II}$ mean value at $z>7$, an additional argument in favour of the use of a constant clumping factor. Similarly, a constant value of the escape fraction is favoured over a redshift-dependent model. When added as a fit parameter, we find $f_\mathrm{esc} = 0.19\pm0.04$. However, this result strongly depends on the choice of magnitude limit in the derivation of $\rho_\mathrm{SFR}$. {Our fiducial analysis considers faint galaxies ($M_\mathrm{lim}= -13$) and the result is a well constrained escape fraction of about 0.2, but when $M_\mathrm{lim}=-17$, the number of galaxies available to reionise the IGM is not sufficient to match the observations, so that much higher values of $f_\mathrm{esc}$, approaching $70\%$, are needed.}}
\keywords{Cosmology: dark ages, reionisation, first stars -- Cosmology: cosmic background radiation -- Galaxies: high-redshift -- Galaxies: evolution -- Galaxies: formation}

\maketitle

\section{Introduction}

Around redshift $z\simeq1090$, during the recombination era, protons paired with free electrons to form neutral atoms: the ionisation level of the intergalactic medium (IGM) fell to $0.0001\, \%$ and remained at this level for several billions of years \citep{peebles_1968,sz_1969,seager_2000}. Nevertheless, observations of the Gunn-Peterson effect \citep{gunn_peterson-1965} in quasar spectra inform us that at $z\sim6$, $99.96 \pm 0.03\, \%$ of the IGM hydrogen atoms are ionised \citep{fan_2006}. What happened in the meantime, during the Epoch of reionisation (EoR), is an essential source of information about the evolution of the Universe, the formation of large cosmic structures and the properties of early galaxies, to cite only a few. Thanks to improved observations of the cosmic microwave background (CMB), luminosity functions of galaxies, damping wings of quasars and Ly-$\alpha$ emissions \citep[e.g.][]{schenker_2013,schroeder_2013,madau_2014,planck_2016}, more and high quality data are available. Now the generally accepted scenario is that first star-forming galaxies reionised neutral regions around them between $z\simeq12$ and $z\simeq6$ and then the ionised regions progressively overlapped \citep[e.g.][]{aghanim_1996,becker_2015} so that IGM neutral hydrogen fraction rapidly decreased until quasars took over to reionise helium from $z\simeq3-4$ \citep{mesinger_2016}.
\\
Yet, some doubts remain about the sources of reionisation: some support the hypothesis that quasars could have led the process \citep{madau_2015,khaire_2016,grazian_2018} but star-forming galaxies are often preferred. For instance, \citet{robertson_2015} argue that they were sufficient to maintain the IGM ionised at $z \sim 7$. The most recent value of the integrated Thomson optical depth, deduced from observations of the CMB, equals $\tau_\mathrm{Planck}=0.058\pm0.012$ and is obtained considering an instantaneous reionisation at $z_\mathrm{reio}=8.8\pm0.9$ ended by $z=6$ \citep{planck_2016}. It is much lower than previous observations by the Wilkinson Microwave Anisotropy Probe (WMAP) $\tau_\mathrm{WMAP}=0.088\pm0.014$ for $z_\mathrm{reio}=10.5\pm1.1$ \citep{hinshaw_2013}. This decrease, according to \citeauthor{robertson_2015}, reduces the need for a significant contribution of high-redshift galaxies and allows them to extrapolate galaxies luminosity functions for $10 \lesssim z \leq 30$. 
\\
Like \citet{robertson_2015}, a number of recently published papers assume redshift-independent values of the escape fraction of ionising photons $f_\mathrm{esc}$ and of the clumping factor $C_\ion{H}{II}$ \citep{bouwens_reionization_2015,ishigaki_2015,greig_2016}, which is a questionable hypothesis. The escape fraction depends on numerous astrophysical parameters and, for this reason, it is often a generalised, global and redshift-independent value that is used, for an order of magnitude of $0.1$. Some simulations give expressions of $f_\mathrm{esc}$ as a function of redshift \citep{haardt_2011,kuhlen_2012} or of various parameters such as halo mass or star formation rate \citep{wise_2014,paardekooper_2015}, but these models are rarely combined with observational constraints, aiming to deduce a certain history of reionisation. The situation is similar for the clumping factor: its evolution with redshift can be considered in simulations through various models \citep[e.g.][]{mellema_2006,pawlik_2009,sobacchi_2014}, but these are rarely compared with observations. We must, however, refer to \citet{price_2016} who constrain parametrised models of the escape fraction $f_\mathrm{esc}\,(z)$ with Thomson optical depth and low multipole E-mode polarisation measurements from \cite{planck_2016}, SDSS BAO data and galaxy observations for $3 \lesssim z \lesssim 10$.
\\
We first describe in Sect. \ref{sec:observables} the observables of the reionisation process we will use throughout the analysis: the cosmic star formation rate density, the ionised fraction of the IGM and the Thomson optical depth, for which observational data is available -- described in Sect. \ref{sec:data}; as well as the two key-parameters of this study, the escape fraction of ionising photons and the clumping factor of IGM ionised hydrogen. Then we look in Sect. \ref{sec:results} for the redshift-evolution we will further consider for the star formation rate (SFR) density, extrapolating luminosity functions at $z\gtrsim10$. Doing this, we study the impact of our observational constraints on $\rho_\mathrm{SFR}$. Investigations are then made on the escape fraction value and on how observations can constrain it: we try several parametrisations out -- a redshift-independent one, where $f_\mathrm{esc}$ is free to vary in $[0.1,0.4]$, and a power-law function of $z$. We proceed the same for $C_\ion{H}{II}$, but this time considering several possible parametrisations of its evolution with redshift, mainly from \citet{iliev_2007} and \citet{pawlik_2009}. We conclude with a discussion of our results in Sect. \ref{sec:discussion}, including a test of different values for the magnitude limit, and a summary in Sect. \ref{sec:conclusions}.\\
Throughout this paper, all cosmological calculations assume the flatness of the Universe and use the Planck cosmological parameters \citep{planck_cosmology_2015}: $h = 0.6774$, $\Omega_\mathrm{m} = 0.309$, $\Omega_\mathrm{b} h^{2} = 0.02230$ and $Y_\mathrm{p} = 0.2453$. Unless otherwise stated, all distances are comoving. 

\section{Observables of reionisation}
\label{sec:observables}

\subsection{Drawing the history of reionisation}
\label{subsec:observables_history}

Clues about the reionisation process can be derived from various observables. Under the assumption that star-forming galaxies provided the majority of the photons which ionised the IGM, the star formation rate density, $\rho_\mathrm{SFR}$, can logically give precious information about the EoR. {Values of SFR density with redshift are deduced from luminosity functions (LF) of star-forming galaxies. LF can be observed down to a certain magnitude, but needs to be extrapolated to consider the contribution of unobserved fainter galaxies. Equation \ref{eq:nion_mag} shows how the comoving ionisation rate $\dot{n}_\mathrm{ion}$ is computed from the LF. }
\begin{equation}
\label{eq:nion_mag}
\begin{split}
\dot{n}_\mathrm{ion} &= \int_{M_\mathrm{lim}}^{\infty} \phi(M_\mathrm{UV})\ f_\mathrm{esc} (M_\mathrm{UV})\ \xi_\mathrm{ion} (M_\mathrm{UV}) \ \mathrm{d}M_\mathrm{UV} \\
&\simeq \  \langle f_\mathrm{esc} \ \xi_\mathrm{ion} \rangle \, \int_{M_\mathrm{lim}}^{\infty} \phi(M_\mathrm{UV}) \ \mathrm{d}M_\mathrm{UV} \\
& \simeq f_\mathrm{esc}\, \xi_\mathrm{ion}\, \rho_\mathrm{SFR}.
\end{split}
\end{equation}
{The final expression directly relates $\rho_\mathrm{SFR}$ to the cosmic reionisation rate $\dot{n}_\mathrm{ion}$, in units of photons per unit time per unit volume, and is the version we will use in our models. We see that the choice of $M_\mathrm{lim}$ is fundamental as it directly impacts the value of $\rho_\mathrm{SFR}$. 
\citet{bouwens_reionization_2015} state that faint galaxies must contribute to the total UV radiation from galaxies but, assuming they do not form efficiently for lower luminosities \citep[see][]{rees_1977, maclow_1999, djikstra_2004}, \citeauthor{robertson_2015} choose to use $M_\mathrm{lim}=-13$ rather than $M_\mathrm{lim}=-17$, a choice we will discuss in this paper.}

Two important parameters are used in Eq. \ref{eq:nion_mag}: $f_\mathrm{esc}$ and $\xi_\mathrm{ion}$. They describe the fact that only a limited amount of the photons produced by star-forming galaxies eventually end up ionising the IGM: first, they need to have sufficient energy -- above the Ly-$\alpha$ limit, and second, they must escape their host galaxy and reach the IGM. The first condition is conveyed by $\xi_\mathrm{ion}$, the quantity of Lyman continuum photons produced per second and per unit SFR for a typical stellar population. According to \citet{robertson_2015}, we take $\xi_\mathrm{ion} = 10^{53.14}$ Lyc photons s$^{-1}$ M$_\odot^{-1}$ yr. The second condition is conveyed by $f_\mathrm{esc}$, the fraction of ionising radiation coming from stellar populations which is not absorbed by dust and neutral hydrogen within the host galaxy and so does contribute to the process. We note that in Eq. \ref{eq:nion_mag} we chose to consider values of $f_\mathrm{esc}$ and $\xi_\mathrm{ion}$ averaged over magnitude, i.e. the effective values.

Aiming to reproduce observations on the star formation history from $z \sim 30 $ to $z \sim 1$, we choose the four-parameter model suggested by \citet{robertson_2015}, updated from \citet[][Sect. 5, Eq. 15]{madau_2014} and described in Eq. \ref{eq:rho_model} below. According to data, $\rho_\mathrm{SFR}(z)$ follows a first rising phase, over $3 \lesssim z \lesssim 15$, which is expressed in our parametrisation by an evolution $\rho_\mathrm{SFR}(z) \propto (1+z)^{b-d}$, up to a peaking point around $z \sim 1.8$ , that is, when the Universe was around $3.6$ Gyr old. It then declines as $\rho_\mathrm{SFR}\,(z) \propto (1+z)^{b}$ until $z=0$. To stay consistent with observations, we set $b>0$ and $b-d<0$. 
\begin{equation}
\label{eq:rho_model}
\rho_\mathrm{SFR}(z) = a \frac{(1+z)^{b}}{1+ \left( \frac{1+z}{c}\right) ^{d}}.
\end{equation}

{In order to put our results in perspective, we consider different values of the magnitude limit for our study and therefore use another parametrisation of the star formation history, suggested by \citet{ishigaki_2015} and designed to reproduce the rapid decrease of $\rho_\mathrm{UV}(z)$ from $z \sim 8$ towards higher redshifts and but not the bump on luminosity density observed around $z \sim 2$}
\begin{equation}
\label{eq:rho_Ishi}
\rho_\mathrm{UV}(z) = \frac{2\, \rho_\mathrm{UV}(z=8)}{10^{a(z-8)}+10^{b(z-8)}}.
\end{equation}
{Here, $\rho_\mathrm{UV}(z=8)$ is a normalisation factor, and $a$ and $b$ characterise the slope of $\rho_\mathrm{UV}(z)$. This model is more adapted to the study of reionisation in itself, as the process is known to end before $z=4$ and so before the star formation bump. However we cannot limit our analysis to this late-redshift model since the former carries more information about the reionisation history and is therefore more interesting when considering a large amount of free parameters. We note that for $M_\mathrm{lim}=-10$ and $M_\mathrm{lim}=-17$, we use $\xi_\mathrm{ion}=10^{25.2} \ \mathrm{erg}^{-1} \, \mathrm{Hz}$, following \citet{ishigaki_2015}.}\\

Other observations can lead to estimations of the fraction of ionised IGM $Q_\ion{H}{II}$, also called filling factor, which relates to the SFR density via Eq. \ref{eq:QHII_diff}. In this equation, the time-related evolution of $Q_\ion{H}{II}$ depends on two contributions: an ionisation source term, proportional to $\dot{n}_{ion}$, and a sink term due to the competition of recombination. $t_\mathrm{rec}$ is the IGM recombination time defined in Eq. \ref{eq:trec_def} and $\langle n_\mathrm{H} \rangle$ is the mean hydrogen number density, defined by $\langle n_\mathrm{H} \rangle = \frac{X_\mathrm{p}\Omega_\mathrm{b}\rho_\mathrm{c}}{m_\mathrm{H}}$, with $\rho_\mathrm{c}$ the critical density of the Universe. 
\begin{equation}
\label{eq:QHII_diff}
\dot{Q}_{H_{II}} = \frac{\dot{n}_\mathrm{ion}}{\langle n_\mathrm{H}\rangle} - \frac{Q_\ion{H}{II}}{t_\mathrm{rec}},
\end{equation}
\begin{equation}
\label{eq:trec_def}
\frac{1}{t_\mathrm{rec}} = C_\ion{H}{II}\ \alpha_\mathrm{B}(T)\ \left( 1+\frac{Y_\mathrm{p}}{4X_\mathrm{p}}\right) \ \langle n_\mathrm{H}\rangle \ (1+z)^{3}.
\end{equation}
In Eq. \ref{eq:trec_def}, $X_\mathrm{p}$ and $Y_\mathrm{p}$ are  the primordial mass fraction of Hydrogen and Helium respectively. $\alpha_\mathrm{B}(T)$ is the case B recombination coefficient at a fiducial IGM temperature of $T = 20\,000$~K, often considered as the mean temperature around a newly ionised atom. This value is consistent with measurements at $z \sim 2 - 4$ \citep{lidz_2010} but has been estimated to $T \lesssim 10^{4}$~K at $z \sim 5-6$ \citep{becker_2011, bolton_2012}. It fluctuates by a factor of between one and two, depending on the spectrum of the sources and on the time passed since reionisation \citep{hui_2003}. Yet, $\alpha_\mathrm{B}$ is expressed as $\alpha_\mathrm{B}(T)\approx 2.6 \times 10^{-13}\, T_{4}^{-0.76}\ \mathrm{cm}^{3}\, \mathrm{s}^{-1}$ with $T_{4}=T/10^{4}\ \mathrm{K}$ \citep{osterbrock_1989}, in other words, it is a weak function of $T$ so that its variations do not affect our results significantly. We note that, rather than case A, we considered case B recombinations in order to exclude recombinations to the ground state and because we consider that ionisations and recombinations are distributed uniformly throughout the IGM, so that each regenerated photon soon encounters another atom to ionise \citep[][Sect. 9.2.1]{loeb_2013}. 
The clumping factor $C_\ion{H}{II}$ expresses how ionised hydrogen nuclei are distributed throughout the IGM. $C_\ion{H}{II}$ and $t_\mathrm{rec}$ are inversly proportional: the more the matter is aggregated in clumps, the easier for ionised atoms to recombine in these very same clumps.\\
To compare with the evolution derived from Eq.~\ref{eq:QHII_diff}, we considered two parametrisations of the time evolution of the filling factor $Q_\ion{H}{II}$, that we will then use to calculate the integrated Thomson optical depth from data.
The first depicts the reionisation process as a step-like and instantaneous transition with a hyperbolic tangent shape (Eq.~\ref{eq:QHII_tgh}). The second is a redshift-asymmetric parametrisation, described in Eq.~\ref{eq:QHII_pow}, inspired by \citet{douspis_2015}. It uses a power-law defined by two parameters i.e. the redshift at which reionisation ends $z_\mathrm{end}$ and the exponent $\alpha$:
\begin{equation}
\label{eq:QHII_tgh}
Q_\ion{H}{II}(z) = \frac{f_\mathrm{e}}{2}\, \left[ 1 + \text{tanh}\left( \frac{y-y_\mathrm{re}}{\delta y}\right) \right],  
\end{equation}
\begin{equation}
\label{eq:QHII_pow}
Q_\ion{H}{II}(z) = 
\left\{ 
        \begin{array}{ll}
                f_\mathrm{e} & \mathrm{for} \: z< z_\mathrm{end},\\
                f_\mathrm{e}\, \left(\frac{z_\mathrm{early}-z}{z_\mathrm{early}-z_\mathrm{end}}\right)^\alpha & \mathrm{for} \: z>z_\mathrm{end}.
        \end{array}
\right.
\end{equation}
where $y\, (z)=(1+z)^{\frac{3}{2}}$, $y_\mathrm{re}=y\, (z=z_\mathrm{re})$ for $z_\mathrm{re}$ the redshift of instantaneous reionisation and $\delta y = \frac{3}{2}\, (1+z)^{\frac{1}{2}}\, \delta z$.  $z_\mathrm{early}$ corresponds to the redshift around which the first emitting sources form, and at which $Q_\ion{H}{II}\,(z)$ is matched to the residual ionised fraction ($\overline{\mathrm{x}}=10^{-4}$). To be consistent with observations, which give $Q_\ion{H}{II}\,(z\leq6.1)\simeq1$ with very low uncertainty \citep{mcgreer_2015}, we choose $z_\mathrm{end}=6.1$. Furthermore, when comparing our findings with the Planck results we set  $z_\mathrm{re}$ at equal to $8.8$, $z_\mathrm{early}=20$, and
also  $\alpha=6.6$ \citep{planck_2016}.\\

Observations of CMB satellites allow us to estimate the Thomson optical depth $\tau$, integrated over the electron column density to the last scattering surface. It expresses the fraction of photons scattered along the line of sight by free electrons and thus is a direct indicator of the global ionisation rate of the IGM. It is related to the two previously described observables $Q_\ion{H}{II}$ and $\rho_\mathrm{SFR}$ via Eq. \ref{eq:tau_def}, where $c$ is the speed of light in vacuum, $\sigma_\mathrm{T}$ the Thomson scattering cross-section, $H(z)$ the Hubble constant and $f_\mathrm{e}$ the number of free electrons per Hydrogen nucleus. We have assumed that Helium is doubly ionised at $z\leq4$ \citep{kuhlen_2012} and thus have $f_\mathrm{e} = 1 + \eta  Y_\mathrm{p} / 4X_\mathrm{p}$ with $\eta = 2$ for $z\leq4$ and $\eta = 1$ for $z > 4$.
\begin{equation}
\label{eq:tau_def}
\tau(z) = c\ \langle n_\mathrm{H}\rangle\ \sigma_\mathrm{T} \int_{0}^{z} f_\mathrm{e}\ \frac{Q_\ion{H}{II}(z')}{H(z')}\ (1+z')^{2}  \ \mathrm{d}z'
\end{equation}

\subsection{Configuring the key-parameters of reionisation}
\label{subsec:observables_parameters}

Among the various parameters cited in Sect. \ref{subsec:observables_history}, two key-parameters of the reionisation history are still under a lot of investigations: the escape fraction and the clumping factor. As mentioned before, $f_\mathrm{esc}$ expresses the fraction of the ionising radiation produced by stellar populations which is not absorbed by dust and neutral hydrogen within its host galaxy, and thus contributes to the ionisation of the IGM. In our approach, it is an effective value, averaged  over stochasticity, halo mass dependencies in the source populations and, most importantly, over all sources considered in the Universe. This averaged value is hard to compare with observations of lone galaxies or haloes, which usually give much lower values. 
For instance, \citet{steidel_2001} and \citet{iwata_2009} estimate the escape fraction of some  $z\sim3$ galaxies to be $\gtrsim 1\%$. On the contrary, overall values  of $f_\mathrm{esc}$ can be derived from simulations but are still highly uncertain.
According to \citet{finkelstein_2015} and to agree with Ly-$\alpha$ forests measurements \citep{bolton_2007}, it should not be higher than $ 0.13 $; \citet{fernandez_2013} use a value of $0.1$ from a simulation; \citet{robertson_2015} deduce from their analysis that, in order to have star-forming galaxies driving the reionisation process at high redshift, $f_\mathrm{esc}$ must equal at least $0.2$; \citet{inoue_2006} find that, if recent values of the escape fraction can be as low as $f_\mathrm{esc}=0.01$ at $z \sim 1$, $f_\mathrm{esc}$ increases quickly with redshift to reach $10\%$ at $z \gtrsim 4$. Finally, \citet{dunlop_2013} assure that, considering the spectral energy distributions observed from high-redshift galaxies, it should be $\approx 0.1 - 0.2$. \citet{yoshiura_2016} summarise results on $f_\mathrm{esc}$ by saying that if it is generally acknowledged that, among all dependencies, the escape fraction decreases with the mass of the galaxy, there is a variance within one or two orders of magnitude among simulations results. For instance, a simulation from \citet{yajima_2014}, on which assumptions of \citet{robertson_2015} are based, shows that, amidst all types of photons produced in star-forming galaxies (Ly-$\alpha$, UV-continuum and ionising photons), the escape fraction of ionising photons is the only one which seems to depend neither on the redshift nor on the galaxy properties: it keeps a constant value of $0.2$ with time, that we use for our first analysis. 

However, photons from different ranges of energy are subject to different physical phenomena and thus escape more or less easily from their host galaxy. For instance, dust extinguishes ionising, Ly-$\alpha$ and UV continuum photons similarly, but only ionising photons are also absorbed by neutral hydrogen clumps. Thus, at high redshifts, when there is little dust around the galaxy, photons of all energy ranges escape as easily; on the contrary, at low redshift, ionising photons experience more difficulties to escape than others \citep{yajima_2014}. We can then infer an increase of $f_\mathrm{esc}$ with redshift that we parametrise in Eq. \ref{eq:fesc_vs_z}, defined for $z\geq4$ and inspired by \citet{kuhlen_2012,chisholm_2018}. This evolution corresponds to either an evolution of the SFR of galaxies themselves and its associated feedback, or by a redshift evolution in the make up of the galaxy population. Here, owing to the UV spectral slope constraints, we set a maximum of $1$ for the $f_\mathrm{esc}$ value, corresponding to a situation where all ionising photons escape. 
\begin{equation}
\label{eq:fesc_vs_z}
f_\mathrm{esc}(z) = \alpha \left( \frac{1+z}{5}\right)^{\beta}  
.\end{equation}
In this parametrisation, also close to the one used in \citet{price_2016}, $\alpha$ is the value of $f_\mathrm{esc}$ at $z=4$ and $\alpha \beta / 5$ of its derivative at $z=4$, redshift at which we expect the hydrogen ionising background to be dominated by star-forming galaxies \citep{kuhlen_2012}. We take $\beta$ positive in order to have an increasing escape with redshift, as anticipated earlier.\\ 

The second key-parameter of the reionisation process which we are going to investigate is the clumping factor of ionised hydrogen in the IGM $C_\ion{H}{II}$, used in Eq. \ref{eq:trec_def}. It expresses how ionised hydrogen nuclei are gathered in heaps throughout the IGM. This parameter is essential because it is the growth of these clumps that allows the reionisation front to progress in the IGM and because competing recombinations will predominantly take place there. A precise estimate of $C_\ion{H}{II}$ can be difficult to obtain. Simulations do indeed have several obstacles to overcome: getting a sufficient precision for the gas distribution, a correct topology of ionised and neutral matter, and  an accurate model of the evolution of gas clumps themselves during the reionisation process. Besides, $C_\ion{H}{II}$ is often first defined on a single ionisation bubble and then summed on all bubbles to get the global volume-averaged value used here: the simulation must consider an extremely wide range of scales \citep[][Sec. 9.2]{loeb_2013}. 

Most recent studies use values ranging from one to six at the redshifts of interest, i.e. for $6 \lesssim z \lesssim 30$  \citep{sokasian_2003,iliev_2006,raicevic_2011,shull_2012,robertson_2015,finkelstein_2015, bouwens_reionization_2015}. Other studies predict a redshift-dependent evolution \citep{iliev_2007,pawlik_2009, haardt_2011, finlator_2012, sobacchi_2014}, justified by the fact that during the late stages of EoR, ionisation fronts penetrate into increasingly overdense regions of the IGM, which have higher recombination rates and so drive a rapid increase of $C_\ion{H}{II}$ \citep{furlanetto_2005,sobacchi_2014}. 
In our study, besides constant values of $C_\ion{H}{II}$, we  consider two parametrisations son the redshift range $3 \leq z \leq 30$\footnote{We assume that $C_\ion{H}{II}$ is the same for \ion{H}{II} and \ion{He}{III} on this range.} :
\begin{equation}
\label{eq:ch2_model_hm}
C_\ion{H}{II}(z) = \alpha + a\, \left(\frac{z}{8}\right)^{b},
\end{equation}
\begin{equation}
\label{eq:ch2_model_iliev}
C_\ion{H}{II}(z) = a\, \mathrm{e}^{\,b\,(z-8)\, +\, c\,(z-8)^{2}}.
\end{equation}
The first expression comes from \citet{haardt_2011}. We update it in order to have $a=C_\ion{H}{II}\, (z=8)- \alpha$ because $Q_\ion{H}{II}$ is close to 0.5 at $z=8$. The second one comes from \citet{mellema_2006} and \citet{iliev_2007} and shows a different behaviour: it is convex and has a minimum at $z_\mathrm{min}=-b/2c$. As explained earlier, it is generally admitted that the clumping factor only decreases with $z$, and therefore we set $z_\mathrm{min} \gtrsim 30$ so that $C_\ion{H}{II}$ does not reach its minimum on our analysis range. For the same reason, $a$ and $b$ from Eq. \ref{eq:ch2_model_hm} have to be of opposite signs and more precisely we take $a>0$ and $b<0$ in order to have $C_\ion{H}{II}(z) \underset{z\rightarrow0}{\longrightarrow} +\infty$. 

{The formal definition of the clumping factor is \citep{bouwens_reionization_2015,robertson_2015}:
$ C_\ion{H}{II} = \langle n_\ion{H}{II} ^2 \rangle / \langle n_\ion{H}{II} \rangle ^2 = 1 + \delta_\ion{H}{II}$,
if we define the overdensity of ionised Hydrogen as $\delta_\ion{H}{II} = \left( n_\ion{H}{II} - \langle n_\ion{H}{II} \rangle \right) / \langle n_\ion{H}{II} \rangle $. Long before the EoR, most of the Hydrogen was neutral so that fluctuations in the ionised Hydrogen overdensity were very weak. In this perspective, we consider in our models that $\delta_\ion{H}{II} (z \rightarrow \infty) = 0$ and so take $C_\ion{H}{II} (z=100) = 1$.}

\section{Data}
\label{sec:data}

The SFR density can be estimated via the observed infrared and rest-frame UV LFs. We use the luminosity densities and SFR densities compiled by \citet{robertson_2015}, computed from \citeauthor{madau_2014} (2014), \citet{schenker_2013}, \citet{mclure_2013}, \citet{oesch_2015} and \citet{bouwens_reionization_2015}. \citeauthor{robertson_2015} also use HST Frontier Fields LF constraints at $z \sim 7$ by \citet{atek_2015} and at $z \sim 9$ by \citet{mcleod_2015}. Estimates of \citet{madau_2014} derived from \citet{bouwens_2012} are updated with newer measurements by \cite{bouwens_reionization_2015}. {For the calculation of $\rho_\mathrm{SFR}$, as a start, luminosity functions of star-forming galaxies are extended to UV absolute magnitudes of $M_\mathrm{lim} = -13$. Then we compared this with results for minimal and maximal magnitude limits $M_\mathrm{lim} = -17$ and $M_\mathrm{lim} = -10$. We note that if \citet{robertson_2015} express $\rho_\mathrm{SFR}$ in M$_\odot$ yr$^{-1}$ Mpc$^{-3}$, \citet{ishigaki_2015} use UV luminosity units, i.e. ergs s$^{-1}$ Hz$^{-1}$ Mpc$^{-3}$. In order to compare results, we used the conversion factor used in \citet{bouwens_reionization_2015} and first derived by \citet{madau_1998}:}
\[
\mathrm{L}_\mathrm{UV} = \frac{\mathrm{SFR}}{\mathrm{M}_\odot \, \mathrm{yr}^{-1}} \times 8.0 \times 10^{27} \ \mathrm{ergs} \,\mathrm{s}^{-1} \, \mathrm{Hz}^{-1}.
\]
{UV luminosity densities used in this work are the ones detailed in \citet{ishigaki_2015}, namely they come from \citet{schenker_2013,mclure_2013,bouwens_2007,bouwens_2014,bouwens_luminosity_2015,oesch_2015}.}
\\
Observations related to the ionised fraction of the IGM $Q_\ion{H}{II}$ used as constraints to our fits include the Gunn-Peterson optical depths and the dark-gap statistics measured in $z \sim 6$ quasars \citep{mcgreer_2015}, damping wings measured in $z \sim 6-6.5$ quasars \citep{schroeder_2013}  and the prevalence of Ly-$\alpha$ emission in $z \sim 7-8$ galaxies \citep{schenker_2013,tilvi_2014,faisst_2014}. We note that in the figures, further data points, not used as constraints in the fit, are displayed to use as comparison. These include observations of Lyman-$\alpha$ emitters \citep{konno_2017,ouchi_2010,ota_2008,caruana_2014}, of near-zone quasars \citep{mortlock_2001,bolton_2011} and of a gamma-ray burst \citep{chornock_2014}.
\\
Last, we consider estimations of the Thomson optical depth derived from Planck Satellite observations: $\tau_\mathrm{Planck} = 0.058 \pm 0.012$ for a redshift of instantaneous reionisation $z_\mathrm{reio} = 8.8 \pm 0.9$ \citep{planck_2016}. We compare it to the asymptotic value $\tau$ obtained from our model calculations at high redshift.

\section{Results}
\label{sec:results}

\subsection{Cosmic star formation history}
\label{subsec:results_SFR}

Since we are interested in the reionisation history both up to and beyond the limit of the current observational data, we adopt the four-parameter model from Eq. \ref{eq:rho_model} into a Monte Carlo Markov chain (MCMC) approach. We  perform a maximum likelihood (ML) determination of the parameter values assuming Gaussian errors on a redshift range of $[0,30]$, extrapolating current observations on star formation history from $z=10.4$ to $z=30$. We fit to the star formation data described in Sect. \ref{sec:data} and then compute the range of credible reionisation histories for every value of the $\rho_\mathrm{SFR}$ model parameters by solving the differential equation of Eq. \ref{eq:QHII_diff}. Filling factor data is used as an additional observational prior for the fit. Finally, we evaluated the Thomson optical depth as a function of $z$ via Eq. \ref{eq:tau_def} and compare its `asymptotic' value, at $z=30$, to $\tau_\mathrm{Planck}=0.058\pm0.012$ \citep{planck_2016} as a last constraint on the fit. Because we want to know what observable constrains reionisation history the most, all constraints are not always used: the run \textbf{ALL} uses all three sets of data as constraints; \textbf{NOQ} skips $Q_\ion{H}{II}$ data; \textbf{NORHO} skips star formation data, and \textbf{ORHO} uses only star formation history in the fit. 

\begin{figure}[t]
\resizebox{\hsize}{!}{\includegraphics{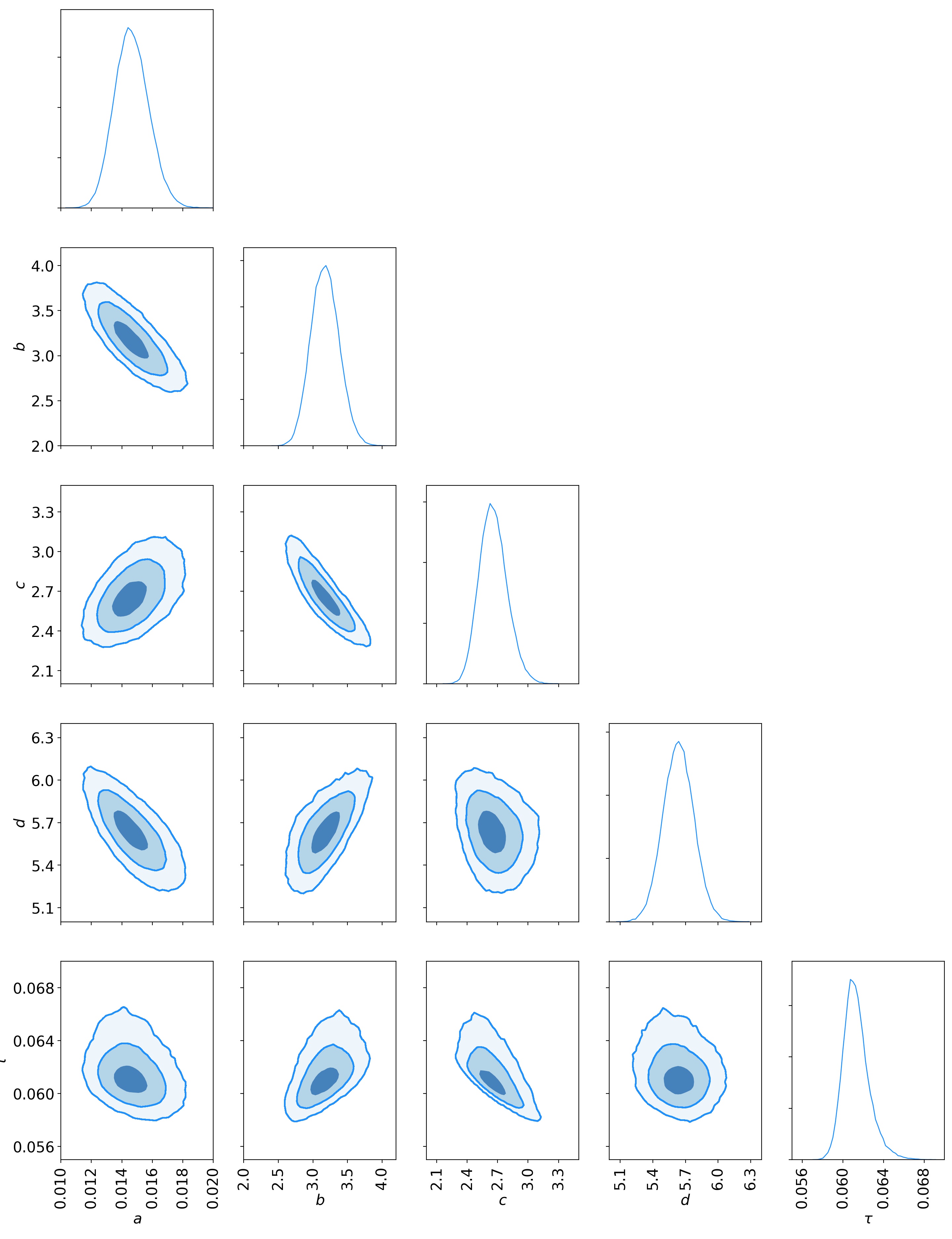}}
\caption{Results of the MCMC analysis for the \textbf{ALL case}. The contours show the 1, 2, 3 $\sigma$ confidence levels for $a, b, c, d$ and the derived parameter $\tau$. }
\label{fig:R_triangle}
\end{figure}

\begin{figure}[h]
\centering
\begin{subfigure}{0.5\textwidth}
\includegraphics[width=1\textwidth]{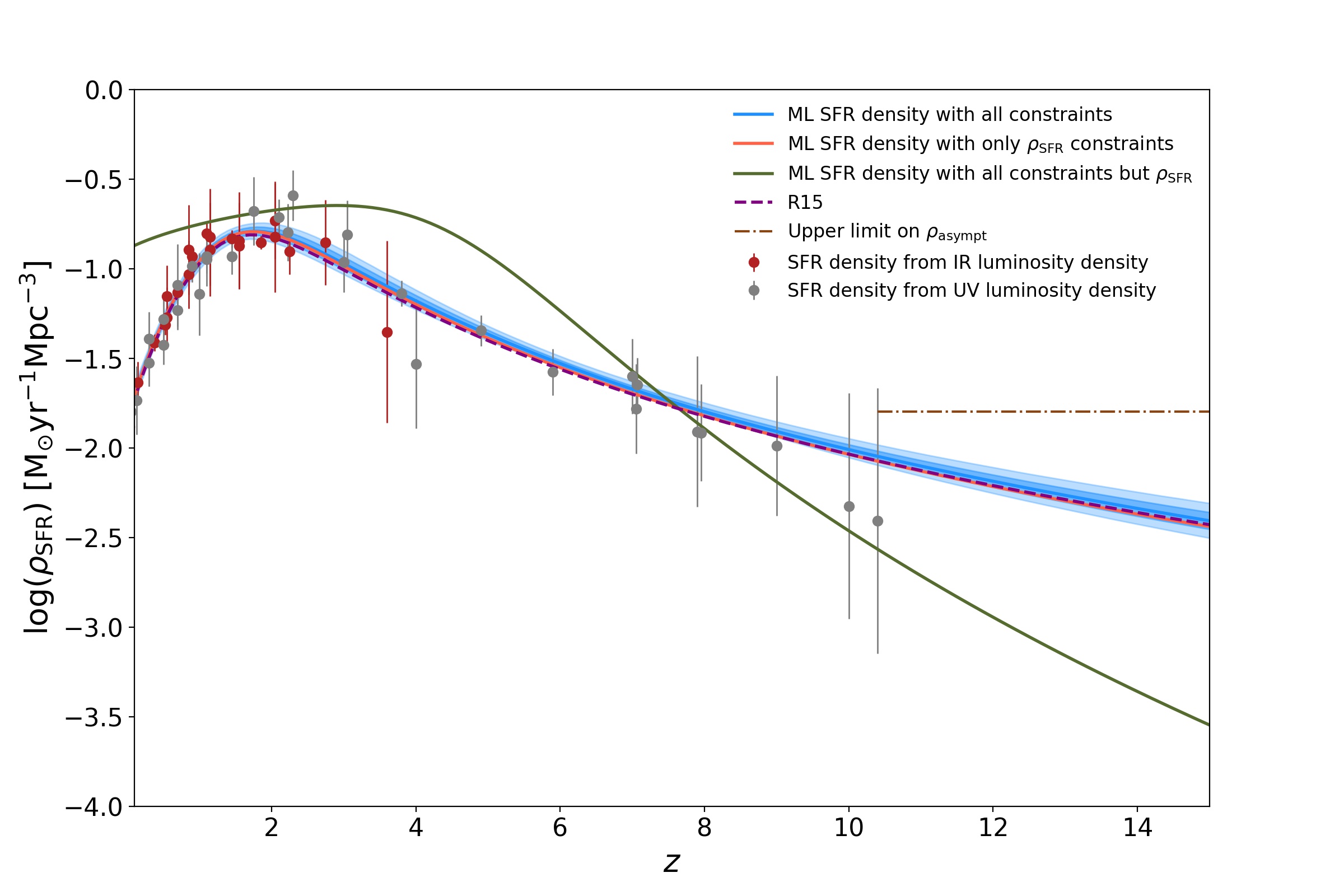}
\caption{} 
\label{fig:SFR_rho_evolution}
\end{subfigure}
\\
\begin{subfigure}{0.5\textwidth}
\includegraphics[width=1\textwidth]{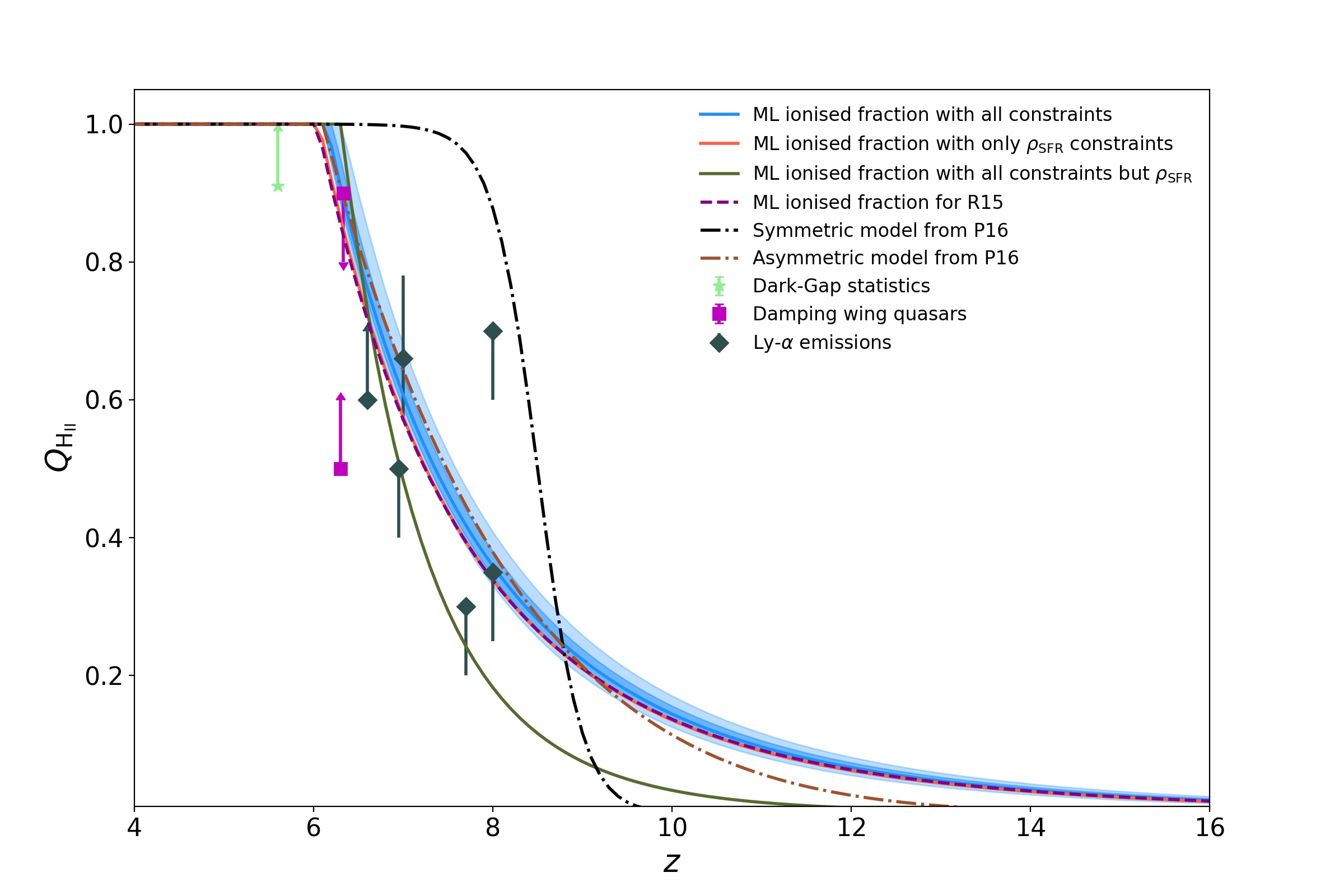}
\caption{}
\label{fig:SFR_QHII_evolution}
\end{subfigure}
\caption{(a) Star formation rate density $\rho_\mathrm{SFR}$ with redshift. Data points are determined from infrared (plotted in red) or ultraviolet (in grey) luminosity densities (Sect. \ref{sec:data}). Maximum likelihood parametrisations (continuous lines) are shown for various set of constraints: blue when all constraints are used; coral when only data on star formation are used; green when $\tau$ and reionisation history data are used. The $68\%$ confidence interval on $\rho_\mathrm{SFR}$ (light blue region) is drawn for the blue model. We note that the interval, corresponding to statistical uncertainties, is very narrow. These inferences are compared with a model forced to reproduce results from \citet{robertson_2015}, cited as R15 in the legend, drawn as the purple dotted line. The horizontal dashed-dotted line corresponds to the upper limit on a hypothetical constant value of $\rho_\mathrm{SFR}$ for $z > 10.4$ (Section \ref{subsec:discussion_sources}). (b) Ionised fraction of the IGM $Q_\ion{H}{II}$ with redshift for same models as (a). Details on the origin of data points are given in the legend and Sect. \ref{sec:data}. Inferences are also compared with the two evolutions used in \citet[][cited as P16]{planck_2016} to model the reionisation process: a redshift-symmetric hyperbolic tangent as the brown dashed-dotted line and a redshift-asymmetric power-law in black.}
\end{figure}

In this first step, we adopt the fiducial, constant with redshift values $f_\mathrm{esc} = 0.2$, $\mathrm{log_{10}}\ \xi_\mathrm{ion} = 53.14$ [Lyc photons $\mathrm{s}^{-1} \mathrm{M}_{\odot}^{-1}$ yr] and $C_\ion{H}{II} = 3$ \citep[e.g.][]{pawlik_2009,shull_2012,robertson_2013, robertson_2015}.
Results are summarised in Fig. \ref{fig:R_triangle} and in Table~\ref{table:SFR_params}. Fig.~\ref{fig:SFR_rho_evolution} shows resulting star formation history and Fig. \ref{fig:SFR_QHII_evolution} resulting reionisation history. We find that star formation history constrains reionisation the most: both figures show that \textbf{ALL} and \textbf{ORHO} runs give similar evolutions and close ML values for $a$, $b$, $c,$ and $d$ (see Table \ref{table:results_rho_parameters}).  We note that our constraints with \textbf{ORHO} and \textbf{ALL} are dominated by the $\rho_\mathrm{SFR}$ data points at a redshifts of approximately five and the fixed functional form assumed for $\rho_\mathrm{SFR}(z)$; they are fully consistent with \cite{robertson_2015}.  On the contrary, for \textbf{NORHO}, the shape of $\rho_\mathrm{SFR}(z)$ is changed and reionisation begins much later, around $z\sim12$ rather than $z \sim 15$ for other runs. \textbf{NORHO} results must be handled carefully as its parameters probability density functions (PDFs) are extremely spread-out; the \textbf{NORHO} line drawn on figures corresponds to the median values of parameters. All we can conclude is that, when star formation history constraints are skipped, there is a much wider range of possible scenarios. 

Interestingly, Fig. \ref{fig:SFR_QHII_evolution} shows that for each run considering star formation history constraints, the process begins as early as $z=15$. This is hardly compatible with WMAP results which stated that, if we consider reionisation as instantaneous, it should occur at $z_\mathrm{reio} \simeq10.5\pm1.1$ \citep{hinshaw_2013} and so cannot begin before $z=12$. Observations also have an influence on the Thomson optical depth values, as \textbf{NORHO} gives a slightly lower value of $\tau$ ($0.053\pm0.003$ compared to $0.061\pm0.001$ for \textbf{ALL}). Yet, all results remain in the 1-$\sigma$ confidence interval of $\tau_\mathrm{Planck}$.

In the rest of the study we used  the \textbf{ALL} run as our definitive parametrisation for $\rho_\mathrm{SFR}$ evolution with redshift: definitive parameters for Eq. \ref{eq:rho_model} are ($a=0.146$, $b=3.17$, $c=2.65$, $d=5.64$) from Table \ref{table:SFR_params}. ML parameters for other runs can be found in Table \ref{table:results_rho_parameters}.

\begin{table}[h]
\caption{ML model parameters for a model using all three sets of constraints.}
\label{table:SFR_params}
\centering
\begin{tabular}{ccccc}
\hline \hline
a & b & c & d & $\tau$ \\ 
\hline
$0.146$ & $3.17$ & $2.65$ & $5.64$ & $0.0612$ \\ 
$ \pm0.001 $ & $ \pm0.20 $ & $ \pm0.14 $ & $ \pm0.141 $ & $ \pm0.0013 $ \\ 
 \hline
\end{tabular} 
\end{table}

\subsection{Escape fraction of ionising photons $f_\mathrm{esc}$}
\label{subsec:results_fesc}

\begin{figure}
\resizebox{\hsize}{!}{\includegraphics{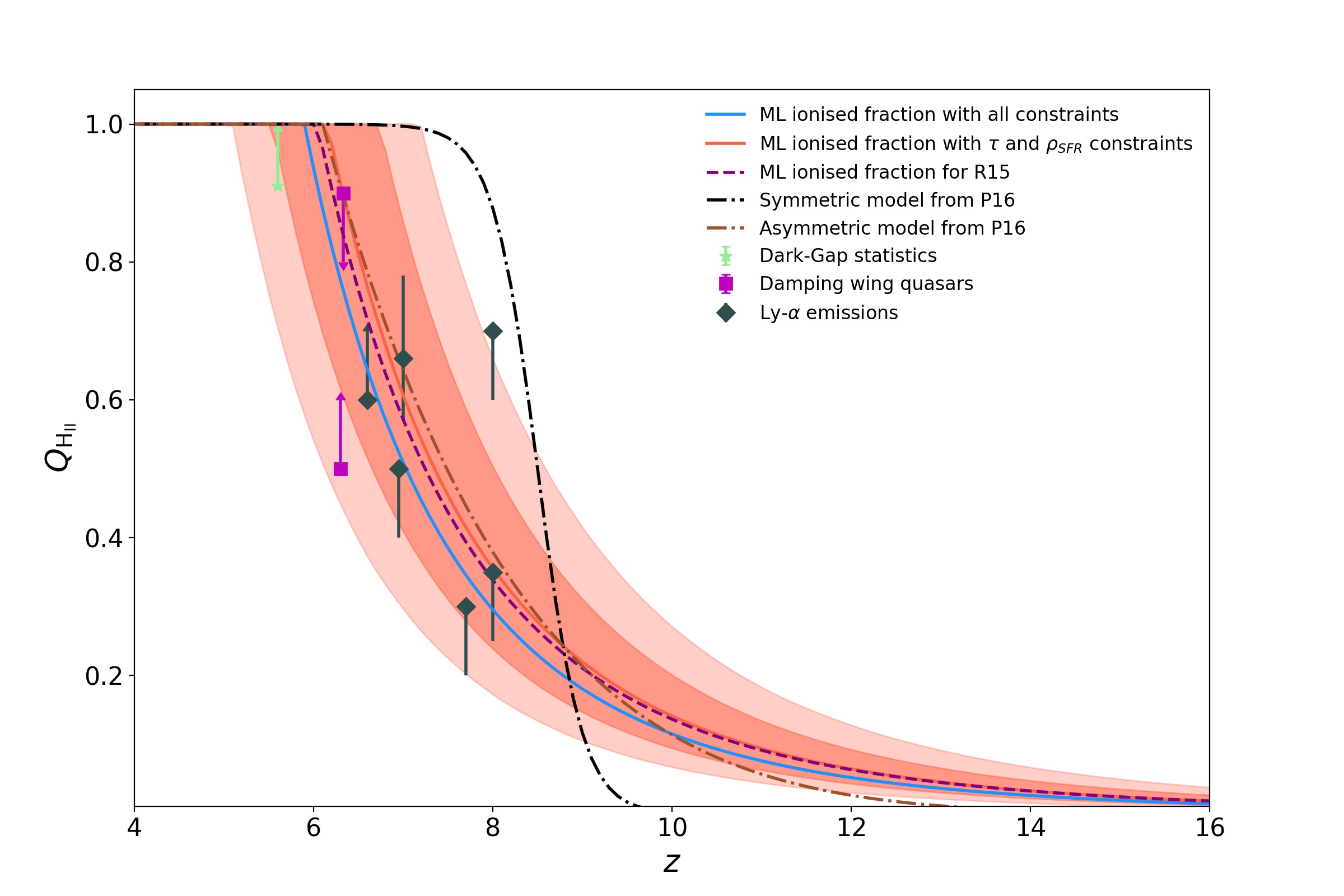}}
\caption{Ionised fraction of the IGM $Q_\ion{H}{II}$ with redshift when $f_\mathrm{esc}$ is introduced as a parameter. Details on the origin of data points are given in the legend. ML models (continuous lines) are shown for various set of constraints: blue when all constraints are used, coral when $Q_\ion{H}{II}$ constraints are skipped, for which the $68\%$ and $95\%$ confidence intervals are drawn in salmon. These inferences are compared with a model forced to reproduce results from \citet[][R15, purple dotted line]{robertson_2015} and with the two evolutions used in \citet[][P16]{planck_2016}: redshift-symmetric as the dashed-dotted brown line and redshift-asymmetric in black.}
\label{fig:fesc_QHII_evolution}
\end{figure}

In order to study the role of the escape fraction in this analysis we chose, as detailed in Sect. \ref{subsec:observables_parameters}, to first consider it as a fifth parameter of the fit -- on top of ($a$, $b$, $c$, and $d$) from Eq. \ref{eq:rho_model}, free to vary between $0$ and $0.4$. We name \textbf{ALL} the run which uses $\rho_\mathrm{SFR}$, $Q_\ion{H}{II}$ and $\tau$ constraints, and \textbf{NOQ} the one that skips ionisation level constraints. $f_\mathrm{esc}$ is involved only in the $\dot{n}_\mathrm{ion}$ calculation of Eq. \ref{eq:nion_mag} and not in the one of $\rho_\mathrm{SFR}$ so that star formation history takes no part in the computation of $f_\mathrm{esc}$. This explains why for all runs, results on the SFR density are close to the ones of Sect. \ref{subsec:results_SFR} (see Tables \ref{table:results_rho_parameters} and \ref{table:results_fesc_ch2_parameters} for details). For \textbf{ALL}, we get ML parameters ($a=0.0147$, $b=3.14$, $c=2.69$, $d=5.74$). Figure \ref{fig:fesc_QHII_evolution} shows that $Q_\ion{H}{II}$ constraints have a strong influence on $f_\mathrm{esc}$: confidence intervals are much wider for \textbf{NOQ} than for \textbf{ALL} (see Table \ref{table:results_fesc_ch2_parameters}). Besides, the \textbf{NOQ} PDF of $f_\mathrm{esc}$ is almost flat: standard deviation is equal to $0.079,$ that is, around $30\%$ of the mean value and two times more than for \textbf{ALL}. For now, we chose to use $f_\mathrm{esc}= 0.19 \pm 0.04$, in other words, the median value of the escape fraction for the \textbf{ALL} run, when a redshift-independent value is needed for $f_\mathrm{esc}$. The full triangle plot for the \textbf{ALL} case is shown Fig.~\ref{fig:Rfesc_triangle} in Appendix. \\

\begin{figure}
\resizebox{\hsize}{!}{\includegraphics{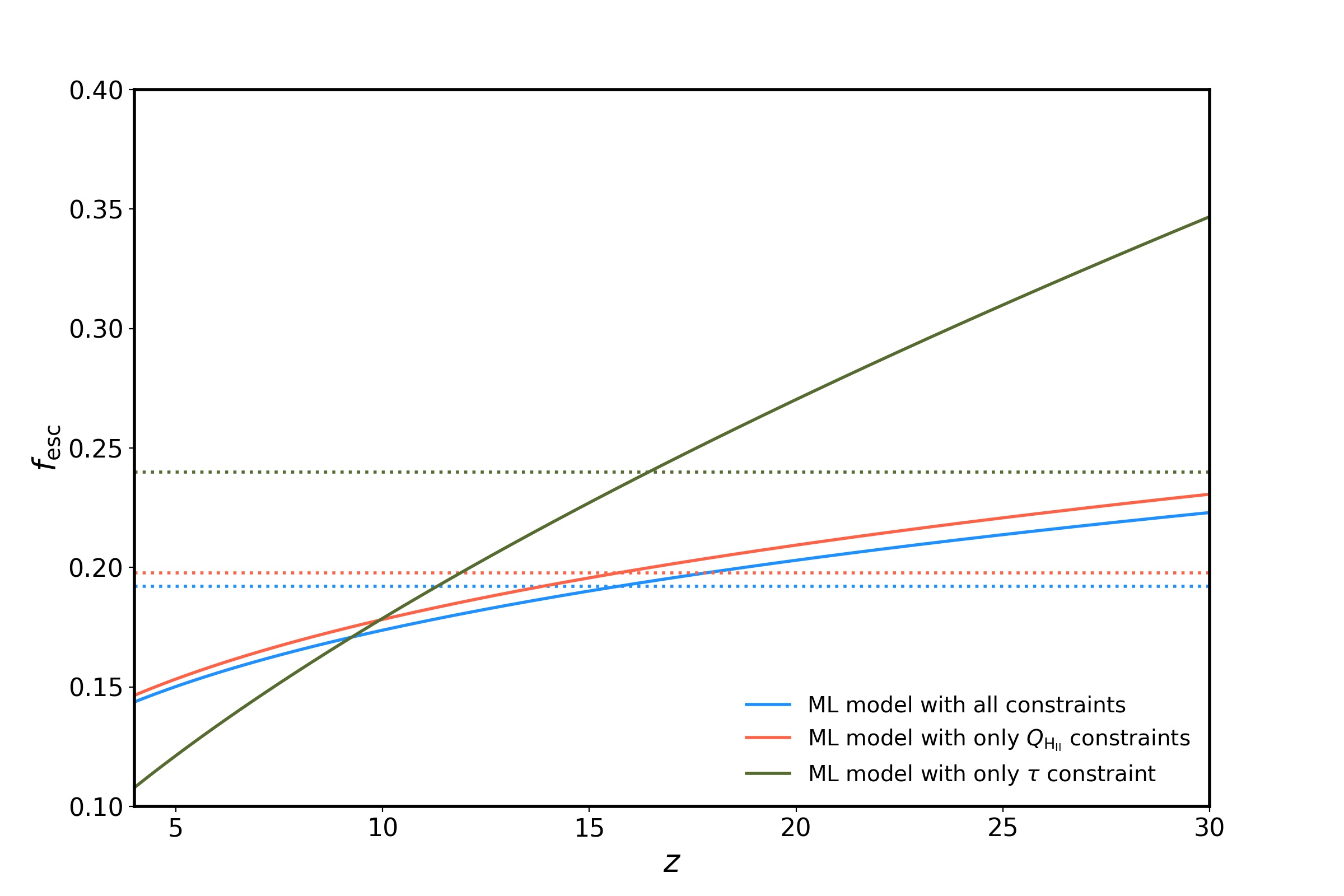}}
\caption{Possible evolutions of $f_\mathrm{esc}$ with redshift. ML models are shown for various set of constraints: blue when all constraints are used; coral when $\tau$ constraints are skipped; green when $Q_\ion{H}{II}$ constraints are skipped. Horizontal dotted lines represent the mean value of $f_\mathrm{esc}$ over $4\leq z \leq30$ for the model of the corresponding colour.}
\label{fig:fesc_xifesc_evolution}
\end{figure}

We now turn to the possibility of a redshift evolution in $f_\mathrm{esc}$ for $z\in [4,30]$. We perform an MCMC maximum likelihood sampling of the two-parameter parametrisation described in Eq. \ref{eq:fesc_vs_z}. For the reasons explained above on the lack of relation between $\rho_\mathrm{SFR}$ and $f_\mathrm{esc}$, we do not use star formation data as a constraint any more and assume that the time evolution of the SFR density follows Eq. \ref{eq:rho_model} using parameters ($a$, $b$, $c$, $d$) resulting from Sect. \ref{subsec:results_SFR}. We used parameters corresponding to the set of constraints that is used on $f_\mathrm{esc}$: if only $\tau$ priors are considered here, we use ($a$, $b$, $c$, and $d$) resulting from a \textbf{NOQ} run (see Table \ref{table:results_rho_parameters} for values).

We find that priors on the IGM ionisation level have a much stronger influence on results than the Thomson optical depth. Indeed, Fig. \ref{fig:fesc_xifesc_evolution} shows that ML evolutions using both $Q_\ion{H}{II}$ and $\tau$ constraints or only $Q_\ion{H}{II}$ are very similar: mean values for $z \geq4$ are similar by $\sim 3\%$ and in both cases, the evolution with redshift is rather weak, as values range from 0.15 around $z \sim 4$ to 0.24 around $z \sim 30$. We note that if \citet{mitra_2015} draw a similar conclusion of an almost constant $f_\mathrm{esc}$ value with redshift from their modelling, they obtain lower values of the escape fraction with an average of about $10\%$ in the redshift range six to nine. For \textbf{NOQ}, the optical depth remains surprisingly close to other models and to $\tau_\mathrm{Planck}=0.058\pm0.012$, around $0.061$. The difference is apparent in the evolution of the ionised fraction, as reionisation begins and ends later, around $z=6$ rather than $z=6.4$ in this case; on the contrary, when $Q_\ion{H}{II}$ data is used, the history tends to be the same as in previous analysis. Our results when only $\tau_\mathrm{Planck}$ constraints are considered are quite similar to those of \citet{price_2016} in which authors study the evolution of $f_\mathrm{esc}$ with redshift. They mainly use constraints from $\tau_\mathrm{Planck}$, concluding to a strong increase of $f_\mathrm{esc}$ from about $0.15$ to about $0.55$, depending on the observational constraints used.
ML parameters for Eq. \ref{eq:fesc_vs_z} when all constraints are considered are ($\alpha=0.14\pm0.02$, $\beta=0\pm0.3$) and give a mean value for $f_\mathrm{esc}$ of about $0.20$, which is extremely close to the $0.19\pm0.04$ found when considering the escape fraction constant with redshift (see Table \ref{table:results_fesc_ch2_parameters} for details).\\
\subsection{Clumping factor of ionised hydrogen in the IGM $C_\ion{H}{II}$}
\label{subsec:results_ch2}

Following the definition of Sect. \ref{subsec:observables_parameters}, we now investigate the constraints on  $C_\ion{H}{II}$ set by observations.
As we did in Sect. \ref{subsec:results_fesc} for $f_\mathrm{esc}$, we added $C_\ion{H}{II}$ as a fifth parameter of the fit on $\rho_\mathrm{SFR}$ using  Eq. \ref{eq:rho_model}, apart from ($a$, $b$, $c$, $d$). It is free to vary between zero and ten, the order of magnitude of fiducial values most commonly used in publications \citep[e.g.][]{shull_2012,robertson_2013, robertson_2015}. Here again, we call \textbf{ALL} the run using all constraints in the fit, and \textbf{NOQ} the one that skips $Q_\ion{H}{II}$ constraints.\\
After performing the MCMC ML sampling of the five parameters (see Table \ref{table:results_rho_parameters} for details), we get a quite spread PDF for $C_\ion{H}{II}$ with \textbf{ALL}: the standard deviation is equal to $1.85$ for a median value of $4.56$. Even with such a wide range of possible values, the range of possible reionisation histories remains very narrow and the Thomson optical depth PDF is almost exactly the same as when we take $C_\ion{H}{II}=3$: $\tau_\mathrm{ALL}=0.0570\pm0.0019$ to be compared with $\tau_{C_\ion{H}{II}=3}=0.0612\pm0.0013$ (see Table \ref{table:ch2_tau} and Fig.~\ref{fig:RCh2_triangle} in Appendix). Besides, for \textbf{NOQ}, the range of possible reionisation histories is wider than for \textbf{ALL}: the width of the $95\%$ confidence area is about $0.6$ when ML reionisation model is halfway through ($Q_\ion{H}{II}=0.5$) for \textbf{NOQ} but $0.16$ for \textbf{ALL}. We also note that for \textbf{NOQ} $\tau$ takes lower values ($\tau_\mathrm{NOQ}=0.0561\pm0.0064$) but remains, as others, in the 1-$\sigma$ confidence interval of $\tau_\mathrm{Planck}$. This confirms that IGM ionisation level data are compatible with Planck observations and that the value of $C_\ion{H}{II}$ constrains only slightly the reionisation history.\\

We now successively test the two redshift-dependent models of the clumping factor given in Eq. \ref{eq:ch2_model_hm} and \ref{eq:ch2_model_iliev}. $C_\ion{H}{II}$ is not involved in the calculation of $\rho_\mathrm{SFR}$ but only of the recombination time. Thus, as for $f_\mathrm{esc}$, star formation history data have no influence over it: the \textbf{ALL} run is now constrained by $Q_\ion{H}{II}$ and $\tau_\mathrm{Planck}$ only. We also  note that, for low values of $z$ (precisely for $z \leq 6.8$), $Q_\ion{H}{II}$ becomes higher that $1$ in our calculations, which is physically irrelevant so we ignore results in this range. 

\begin{figure}
\centering
\begin{subfigure}{0.5\textwidth}
\includegraphics[width=1\textwidth]{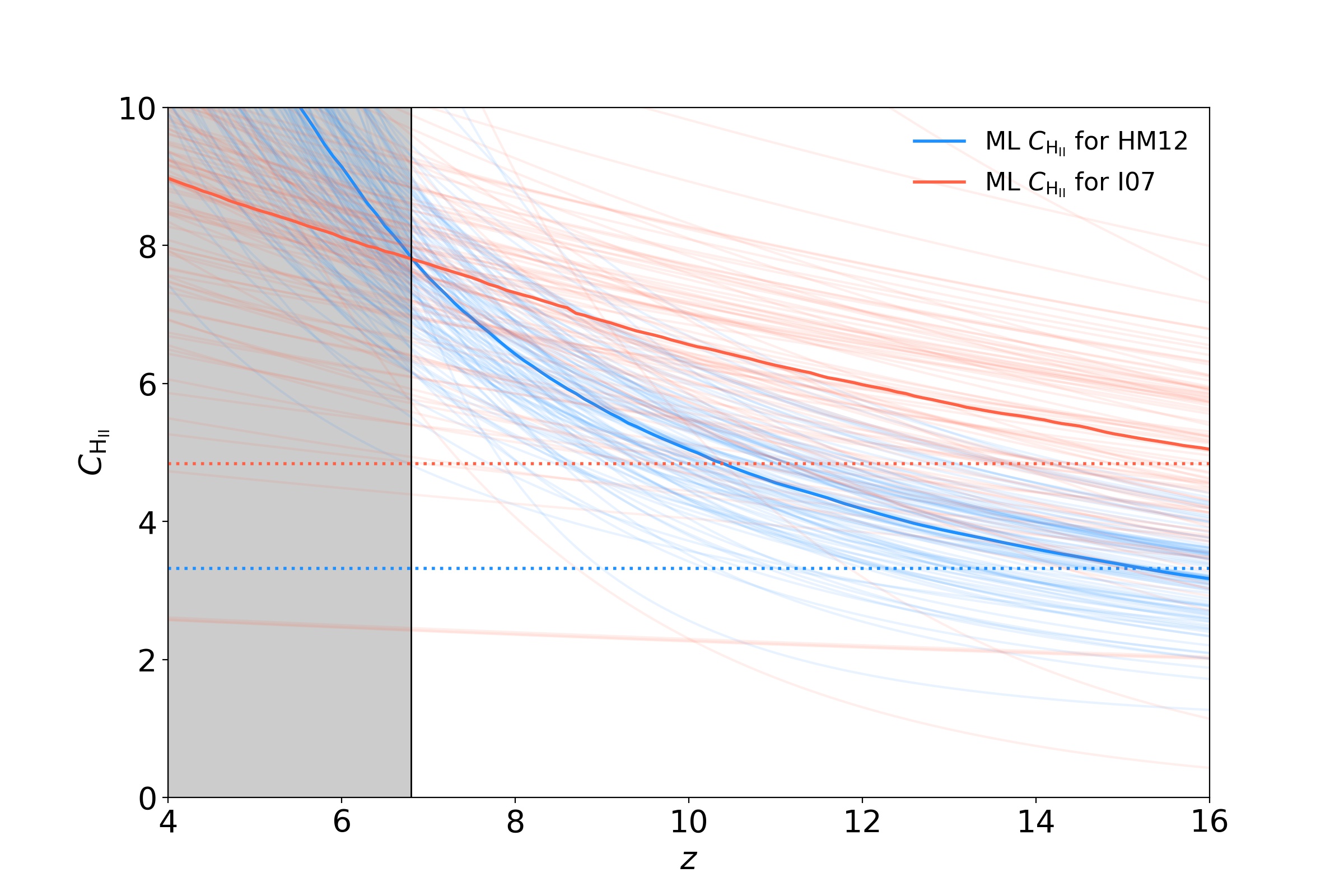}
\caption{} 
\label{fig:ch2_ch2_evolutions}
\end{subfigure}
\\
\begin{subfigure}{0.5\textwidth}
\includegraphics[width=1\textwidth]{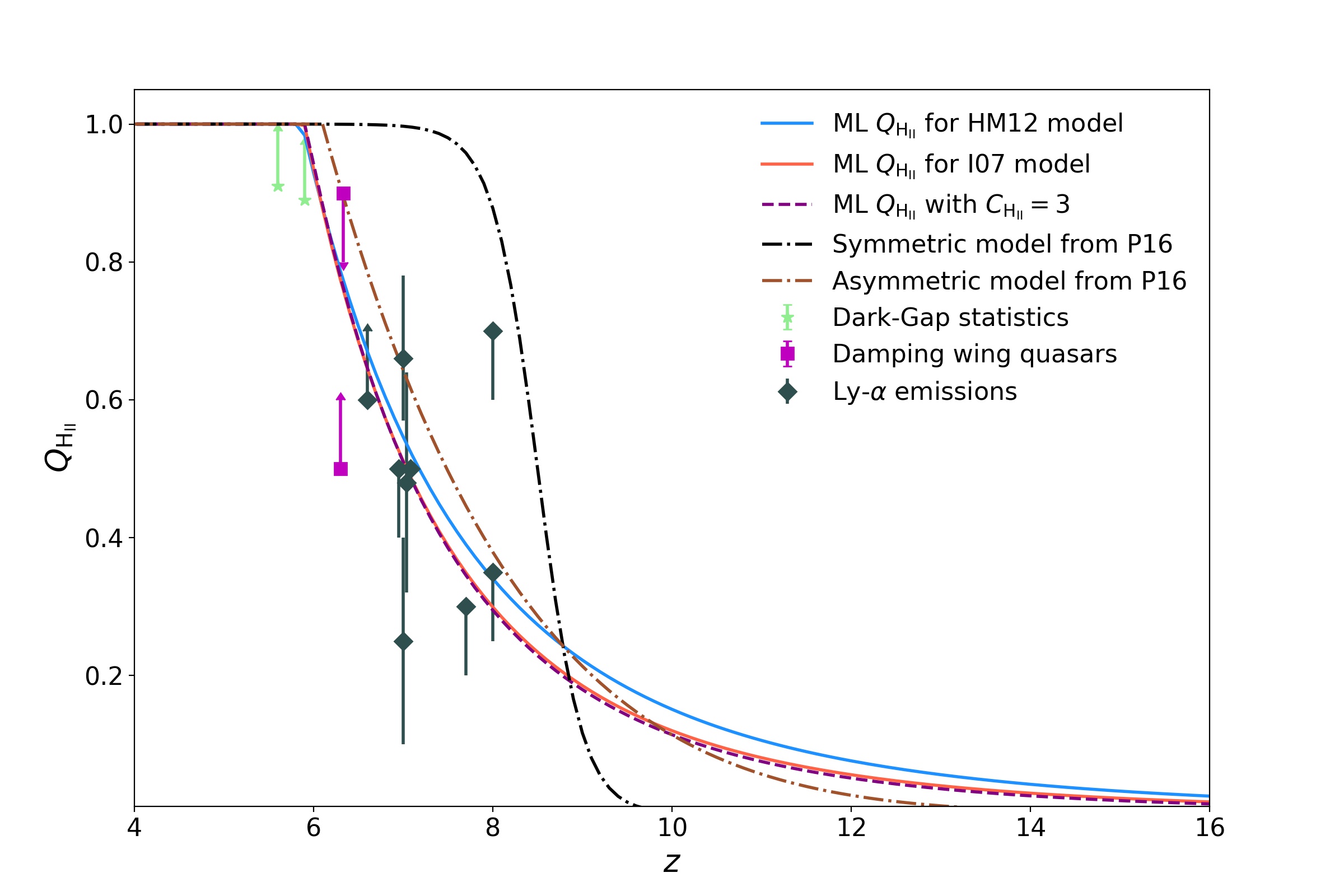}
\caption{}
\label{fig:ch2_QHII_evolutions}
\end{subfigure}
\caption{(a) Possible evolutions of $C_\ion{H}{II}$ with redshift. ML models are shown for the two models of Sect. \ref{subsec:observables_parameters}: blue for the first, coral for the second. Dotted horizontal lines correspond to the mean value of $C_\ion{H}{II}(z)$ for $z>6.8$, where outputs of the model are used in calculations, for the model of the corresponding colour. The vertical line is located at $z=6.8$. Lines of lighter colours represent various outputs of the sampling of the corresponding model. (b) Redshift evolution of $Q_\ion{H}{II}$ for the same models of $C_\ion{H}{II}(z)$. Inferences are compared to a result with $C_\ion{H}{II}(z)=3$ in purple dashed line, and to the theoretical models of \citet{planck_2016}: a redshift-symmetric model in black and a redshift-asymmetric model in brown.\\
\textbf{References.} HM12: \citet{haardt_2011} or Eq. \ref{eq:ch2_model_hm}. I07: \citet{iliev_2007} or Eq. \ref{eq:ch2_model_iliev}.}
\end{figure}

Once again, IGM reionisation level data constrain results more than $\tau_\mathrm{Planck}$. The redshift-evolution of $C_\ion{H}{II}$ and $Q_\ion{H}{II}$ for the two parametrisations presented in Sect. \ref{subsec:observables_parameters} and for \textbf{ALL} runs are shown in Figs. \ref{fig:ch2_ch2_evolutions} and \ref{fig:ch2_QHII_evolutions}. We see on the left panel that there are a lot of possible output evolutions for both models but this does not translate in significant variations of $Q_\ion{H}{II}(z)$ whose $68\%$ confidence intervals are found to be very narrow. All scenarios remain quite close, with reionisation beginning around $z=16$ and ended by $z=6$. This means that, as in previous paragraph where $C_\ion{H}{II}$ was assumed constant with redshift, its exact value has no significant impact on the reionisation history. In fact, variations in $C_\ion{H}{II}$ have some impact on the computed Thomson optical depth: as seen in Table \ref{table:ch2_tau}, higher values of $C_\ion{H}{II}$ allow for a lower value of $\tau$ -- consistent with Eqs. \ref{eq:trec_def} and \ref{eq:tau_def}. All values remains in the 1-$\sigma$ confidence interval of $\tau_\mathrm{Planck}$.\\

Finally, it seems that the fiducial constant value often used in papers,$C_\ion{H}{II}=3$, and which lies between the mean values of our models ($\sim 3$ for \textbf{HM12}, $1.8$ for \textbf{I07}, and $4.5$ for \textbf{Free}), is a reasonable choice. More generally, and in accordance with \citet{bouwens_reionization_2015}, as long it remains in a range of [1.4,8.6], which is the $95\%$ confidence interval of $C_\ion{H}{II}$ from first paragraph (\textbf{Free} fit), results are consistent with the three sets of constraints available. This result corroborates the work of \citet{price_2016}, who also note that their analysis is almost completely independent of the clumping factor over the prior range $1< C_\ion{H}{II} < 5$.

\begin{table}
\centering
\caption{Resulting Thomson optical depths for various evolutions of $C_\ion{H}{II}$ with redshift.}
\label{table:ch2_tau}
\begin{tabular}{ccc}
\hline \hline 
Model & $\langle \tau \rangle$ & $\sigma$ \\ 
\hline
$C_\ion{H}{II}=3$ & $0.0612$ & $0.0013$ \\ 
Free & $0.0570$ & $0.0019$ \\
HM12 & $0.0604$ & $0.0020$ \\ 
I07 & $0.0579$ & $0.0027$  \\
 \hline
\end{tabular} 
\tablebib{Free: Model with $C_\ion{H}{II}$ as a fifth parameter, varying in $[1,10]$; HM12: \citet{haardt_2011}, Eq. \ref{eq:ch2_model_hm}; I07: \citet{iliev_2007}, Eq. \ref{eq:ch2_model_iliev}.
}
\end{table}

\subsection{Varying both $f_\mathrm{esc}$ and $C_\ion{H}{II}$}
\label{subsec:fesc_Ch2}

\begin{figure}
\centering
\resizebox{\hsize}{!}{\includegraphics{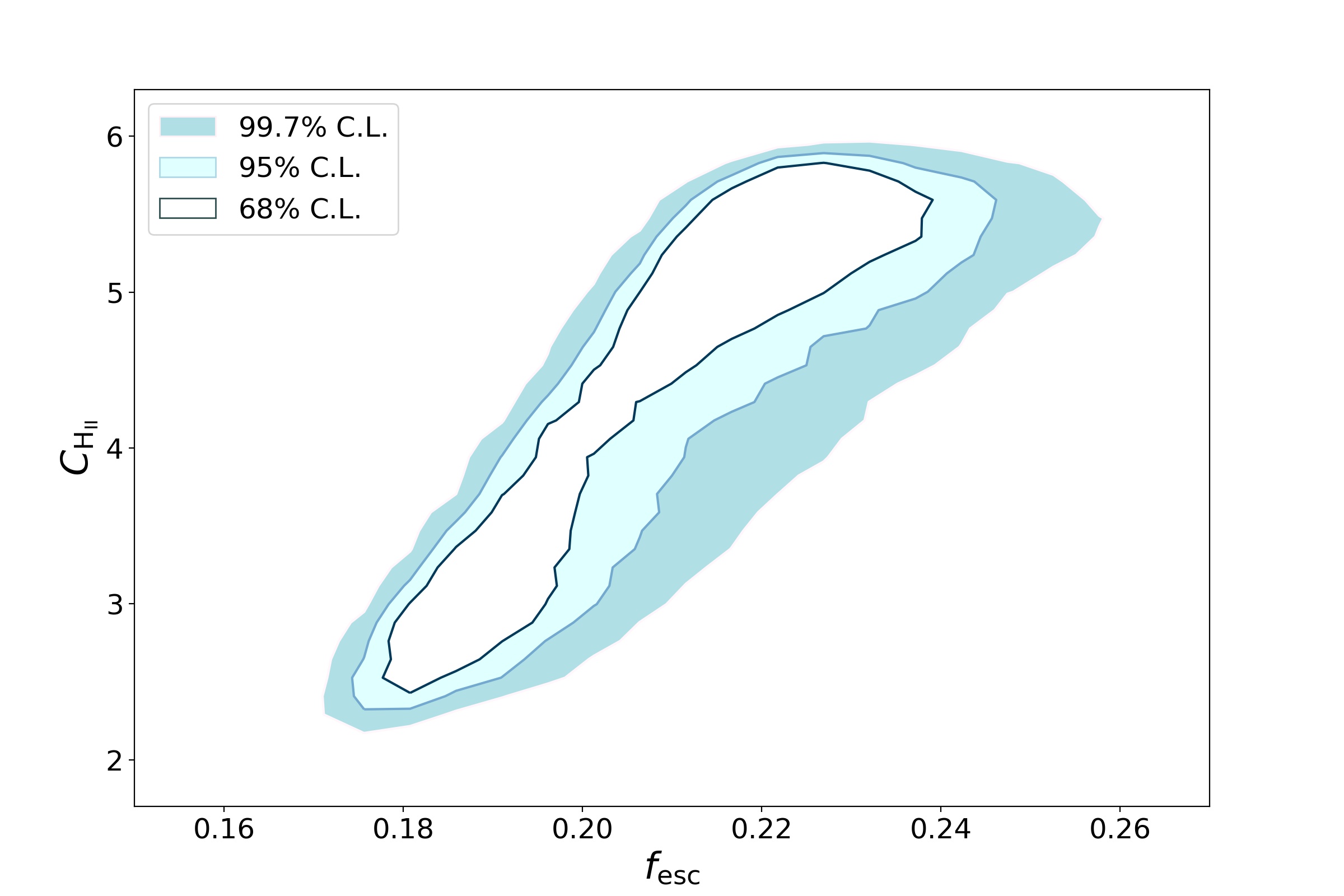}}
\caption{MCMC distribution for $f_{esc}$ and $C_\ion{H}{II}$ when both are taken as fit parameters (other parameters fixed). The escape fraction is allowed to vary between $0.1\%$ and $100\%$, the clumping factor between one and seven.}
\label{fig:ch2_vs_fesc_distri}
\end{figure}

Now we have studied the impact of $f_\mathrm{esc}$ and $C_\ion{H}{II}$ separately, we set the evolution of $\rho_\mathrm{SFR}(z)$ according to Eq. \ref{eq:rho_model}, using parameters $a$, $b$, $c,$ and $d$ resulting from the analysis of Sect. \ref{subsec:results_SFR}. We performed an MCMC maximum likelihood sampling of the two parameters $f_\mathrm{esc}$ and $C_\ion{H}{II}$, considered constant with redshift. The first is allowed to vary between 0.001 and 1, the other between one and seven. We show parameter distributions for $f_{esc}$ and $C_\ion{H}{II}$ in Fig. \ref{fig:ch2_vs_fesc_distri}. We constrain the fit with all three data sets. 

If we consider the median value of each parameter distribution as its maximum likelihood value, we find $f_{esc}=0.193 \pm 0.026$ and $C_\ion{H}{II} = 4.43 \pm 1.11$. We see results are pretty similar to the previous analysis: if the escape fraction is well constrained, with a standard deviation of about $13\%$, the clumping factor can take a much wider range of values, between 3 and 5.5. We note that there seems to be a strong upper bound for the escape fraction around 0.26, which we can compare to the asymptotic value of $f_{esc}$ when it is allowed to change with redshift (see Fig. \ref{fig:fesc_xifesc_evolution}). Because parameters take values close to previous results, the resulting ionisation histories are also close to the ones observed in Fig. \ref{fig:SFR_QHII_evolution} and are hence in good agreement with observations.

\begin{figure}
\centering
\resizebox{\hsize}{!}{\includegraphics{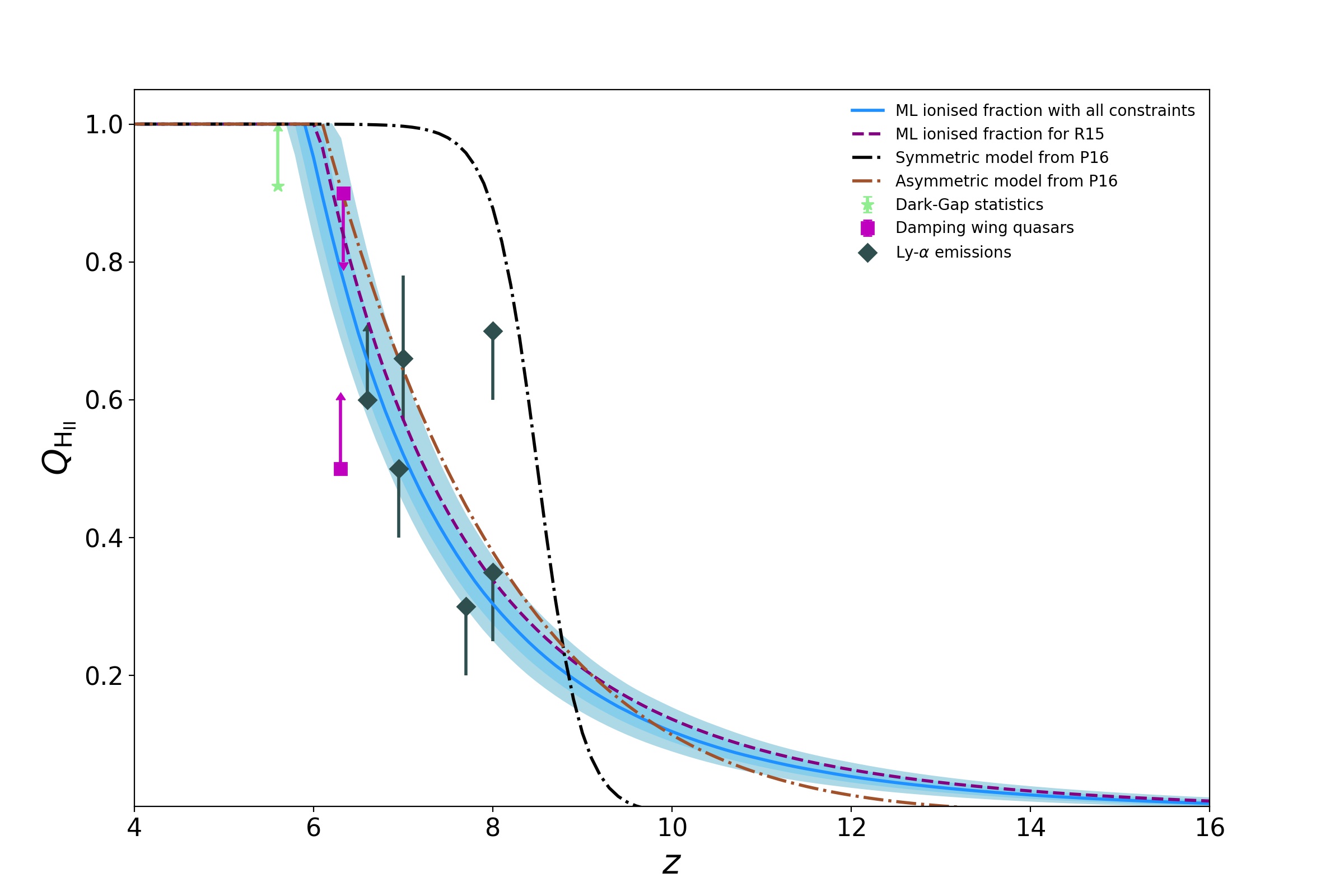}}
\caption{Redshift evolution of $Q_\ion{H}{II}$ when all parameters ($a,b,c,d, f_{esc}, C_\ion{H}{II}$) are free and all datasets used. Fig.~\ref{fig:RfescCh2_triangle} show the corresponding constraints on assumed parameters. }
\label{fig:QH2_all}
\end{figure}

Finally, we considered the case when the four parameters describing
  the evolution of $\rho_\mathrm{SFR}(z)$ are set free in the same
  time as $f_\mathrm{esc}$ and $C_\ion{H}{II}$, using all datasets. We
  assumed the same prior as \cite{price_2016} on $C_\ion{H}{II}$
  considering values between one and five. The full triangle plot is shown
  in Fig.~\ref{fig:RfescCh2_triangle} and best fit parameters are
  reported in Table~\ref{table:results_rho_parameters}. The values
  found are in agreement with previous runs, with an undetermined
  value of $C_\ion{H}{II}$ at the $2\sigma$ level. As in
  \cite{price_2016} the degeneracy between $f_\mathrm{esc}$ and
  $C_\ion{H}{II}$ and the current data do not allow to constrain
  strongly all free parameters. However the evolution of the filling factor (Fig.~\ref{fig:QH2_all}) and thus the derived value of $\tau$ remain quite well constrained ($\tau=0.058\pm0.002$) and in agreement with Planck ($\tau_\mathrm{Planck} = 0.058 \pm 0.012$). 

\section{Discussion}
\label{sec:discussion}

\subsection{Influence of the magnitude limit}
\label{subsec:discussion_Mlim}

In order to study the influence of the choice of magnitude limit on our results, we adopt the model of Eq. \ref{eq:rho_Ishi} into an MCMC approach similar to Sect. \ref{sec:results}. We fit the model to our three data sets adapted to the corresponding magnitude limit as described in Section \ref{sec:data}. $M_\mathrm{lim}=-17$ and $M_\mathrm{lim}=-10$ correspond to the analysis performed in \citet{ishigaki_2015}, and $M_\mathrm{lim}=-13$ corresponds to \citet{robertson_2015}. 

\begin{figure*}[!h]
    \centering
    \begin{subfigure}[b]{0.9\textwidth}
        \centering
        \begin{subfigure}[b]{0.48\textwidth}
        \centering 
        \includegraphics[width=\textwidth]{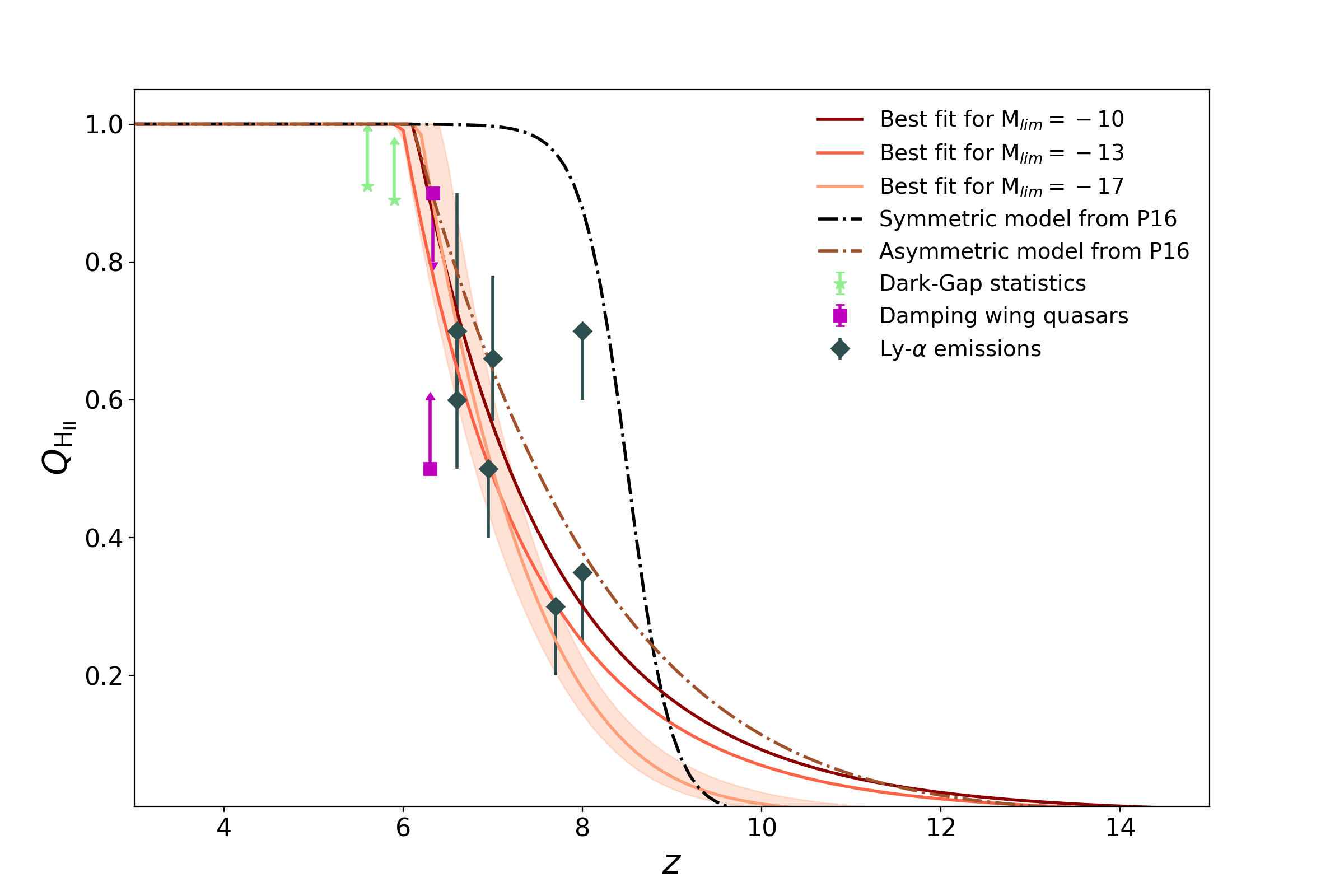}
    \end{subfigure}     
      \quad
    \begin{subfigure}[b]{0.48\textwidth}
    \centering
        \includegraphics[width=\textwidth]{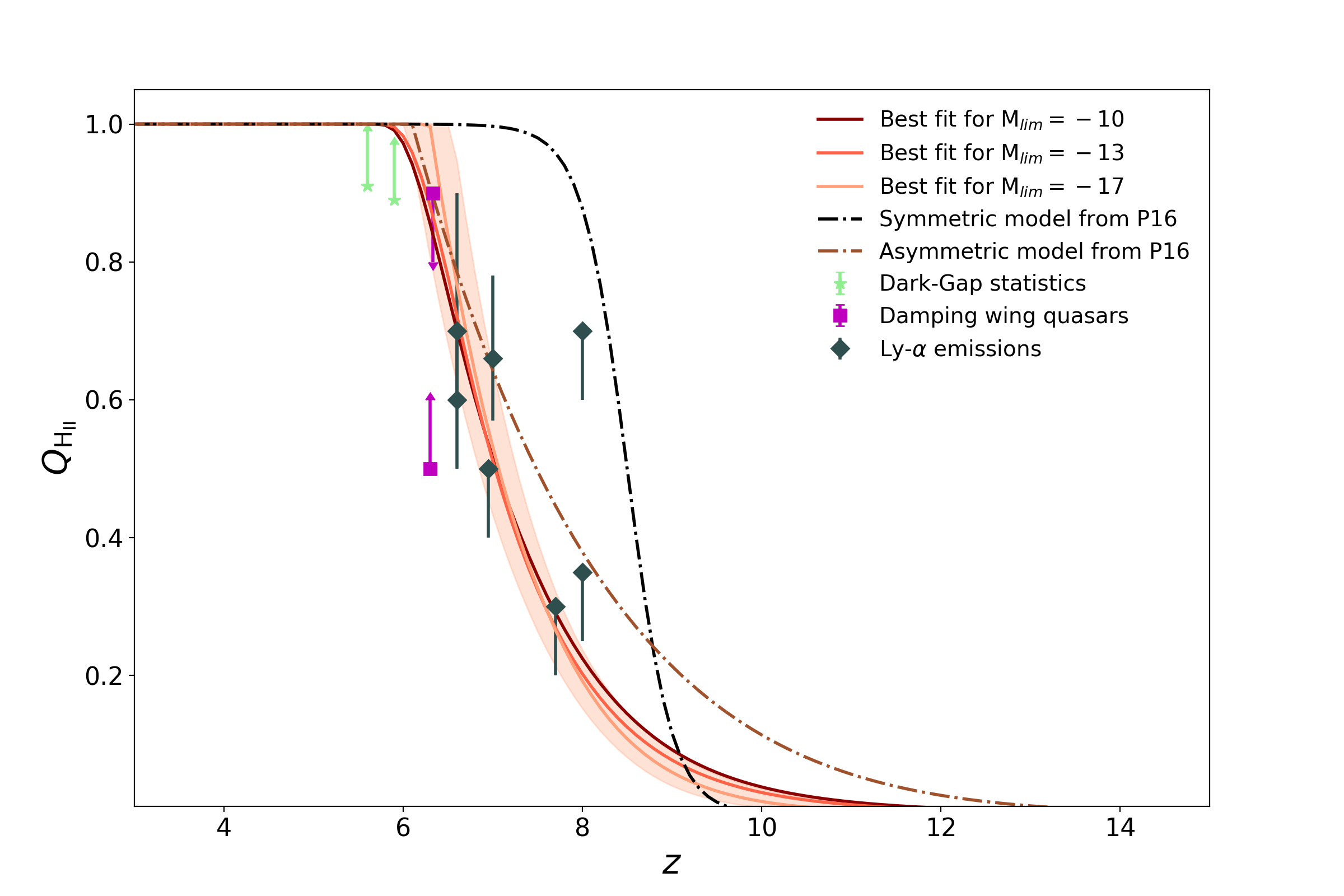}
\end{subfigure}
\caption{}
  \label{fig:mag_comparison_QHII}
  \end{subfigure}
  \vskip\baselineskip
    
    \begin{subfigure}[b]{0.9\textwidth}
    \centering
    \begin{subfigure}[b]{0.475\textwidth}
    \centering
  \includegraphics[width=\textwidth]{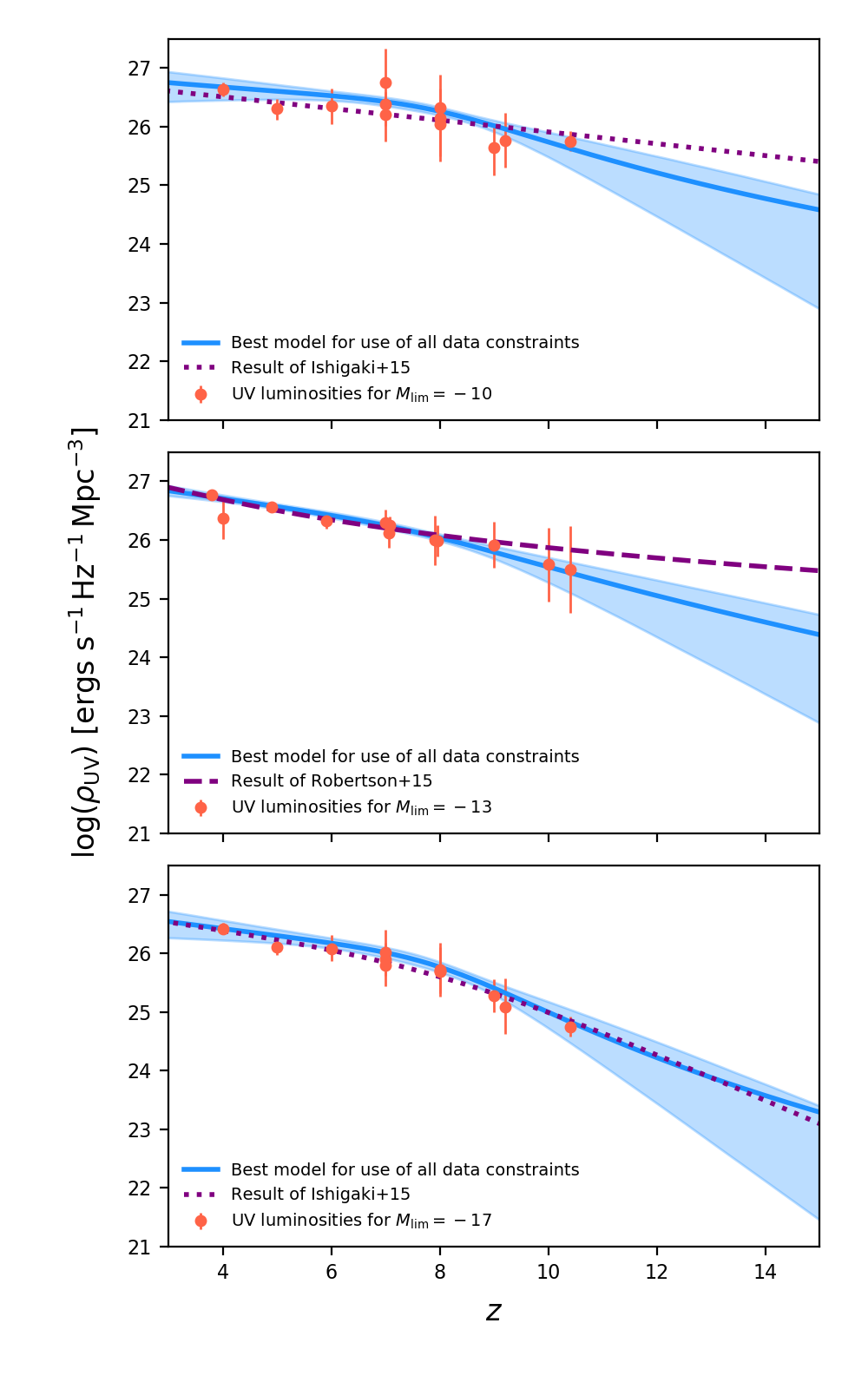}
    \end{subfigure}
    \quad
    \begin{subfigure}[b]{0.475\textwidth}
    \centering
  \includegraphics[width=\textwidth]{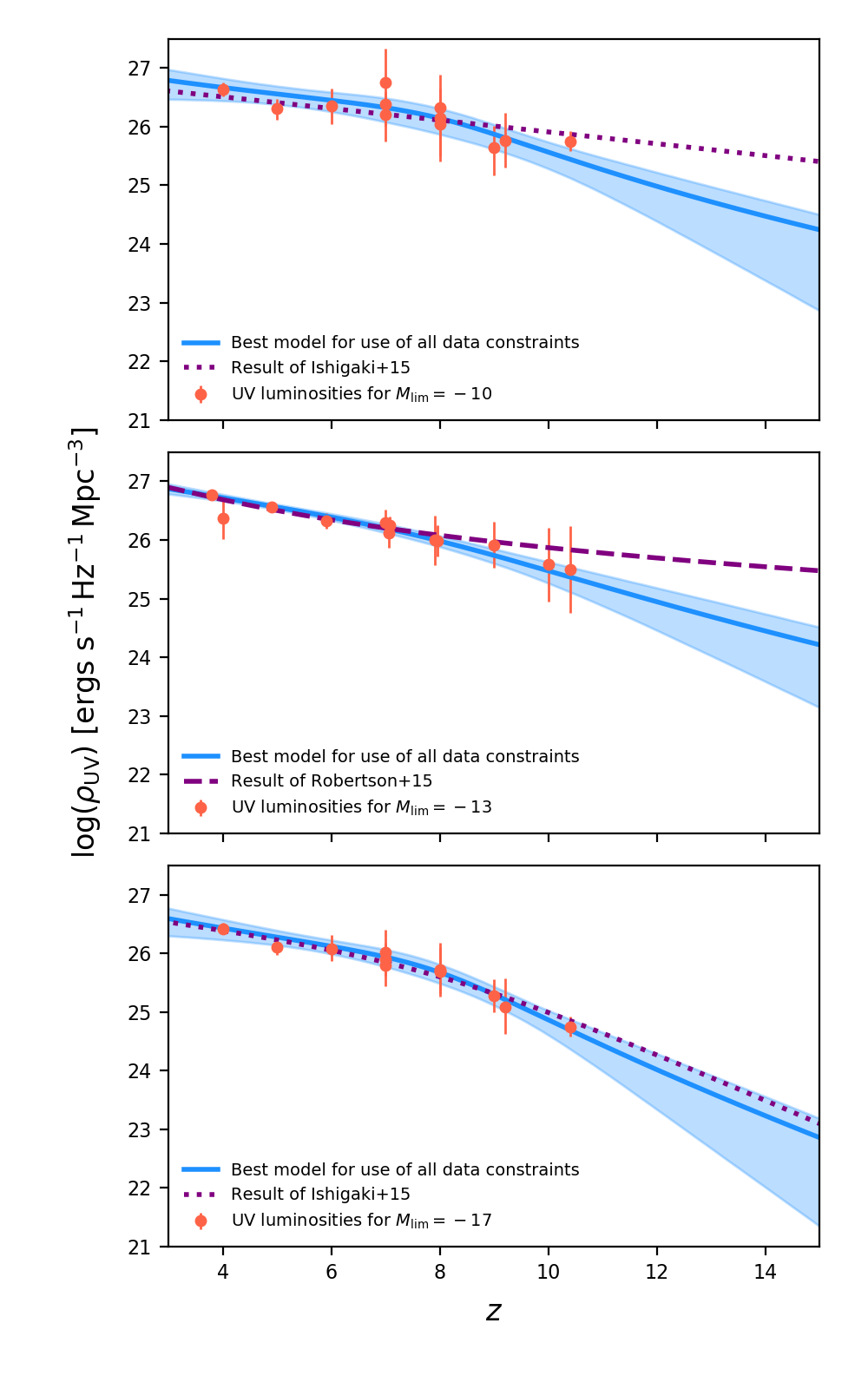}
    \end{subfigure}
\caption{}
\label{fig:mag_comparison_rhoSFR}
\end{subfigure}
\caption{(a) Redshift evolution of $Q_\ion{H}{II}$ for various choices of the magnitude limit in luminosity data: brown for $M_\mathrm{lim}=-10$, orange for $M_\mathrm{lim}=-13$ and beige for $M_\mathrm{lim}=-17$. The light orange region represents the $68\%$ confidence level for the worst case scenario, i.e. $M_\mathrm{lim}=-17$. Left panel: escape fraction fixed to the values used by corresponding references. Right panel:  escape fraction  varying between zero and one. (b) UV luminosity density $\rho_\mathrm{UV}$ with redshift in logarithmic scale for three values of the magnitude limit: $M_\mathrm{lim}=-10$ in the upper panel, $M_\mathrm{lim}=-13$ in the middle panel and $M_\mathrm{lim}=-17$ in the lower panel. Data points are from \citet{ishigaki_2015} or adapted from \citet{robertson_2015}. Maximum likelihood parametrisations corresponding to Eq. \ref{eq:rho_Ishi} (continuous lines) are shown for fits using all observational constraints. The $68\%$ confidence interval is represented as the light blue region. These results are compared with a model forced to reproduce results from corresponding references, drawn as the purple lines. Left panel: escape fraction fixed. Right panel:  escape fraction allowed to vary between zero and one.}
\end{figure*}

We compute the star formation and reionisation histories compatible with the three sets of observational data, for the maximum likelihood parameters (here, median values) of the parametrisation in Eq. \ref{eq:rho_Ishi} and for the three $M_\mathrm{lim}$ values. Results can be found in Figs. \ref{fig:mag_comparison_QHII} and \ref{fig:mag_comparison_rhoSFR} where two cases have been considered: $f_\mathrm{esc}$ fixed, taken to have the value used in corresponding references (left panels) and $f_\mathrm{esc}$ allowed to vary between 0 and 1 (right panels). In both cases, the effect of the two additional sets of data used as constraints here, $Q_\ion{H}{II}$ and $\tau_\mathrm{Planck}$, which were not used in \citet{ishigaki_2015}, is to lower the quantity of ionising sources needed at high redshift to reach a fully ionised IGM by $z \sim 6$.  We note, however, that here the values of some parameters were taken from \citet{ishigaki_2015} and hence quite different from the ones used in Sect. \ref{subsec:results_SFR}. For instance, \citeauthor{ishigaki_2015} found $C_\ion{H}{II}$ values of $1.9$ and $1.0$ for respectively $M_\mathrm{lim}=-17$ and $M_\mathrm{lim}=-10$ whereas we used $C_\ion{H}{II}=3$ before and consequently in the analysis for $M_\mathrm{lim}=-13$.

\begin{figure*}[!h]
\centering
\includegraphics[width=0.7\textwidth]{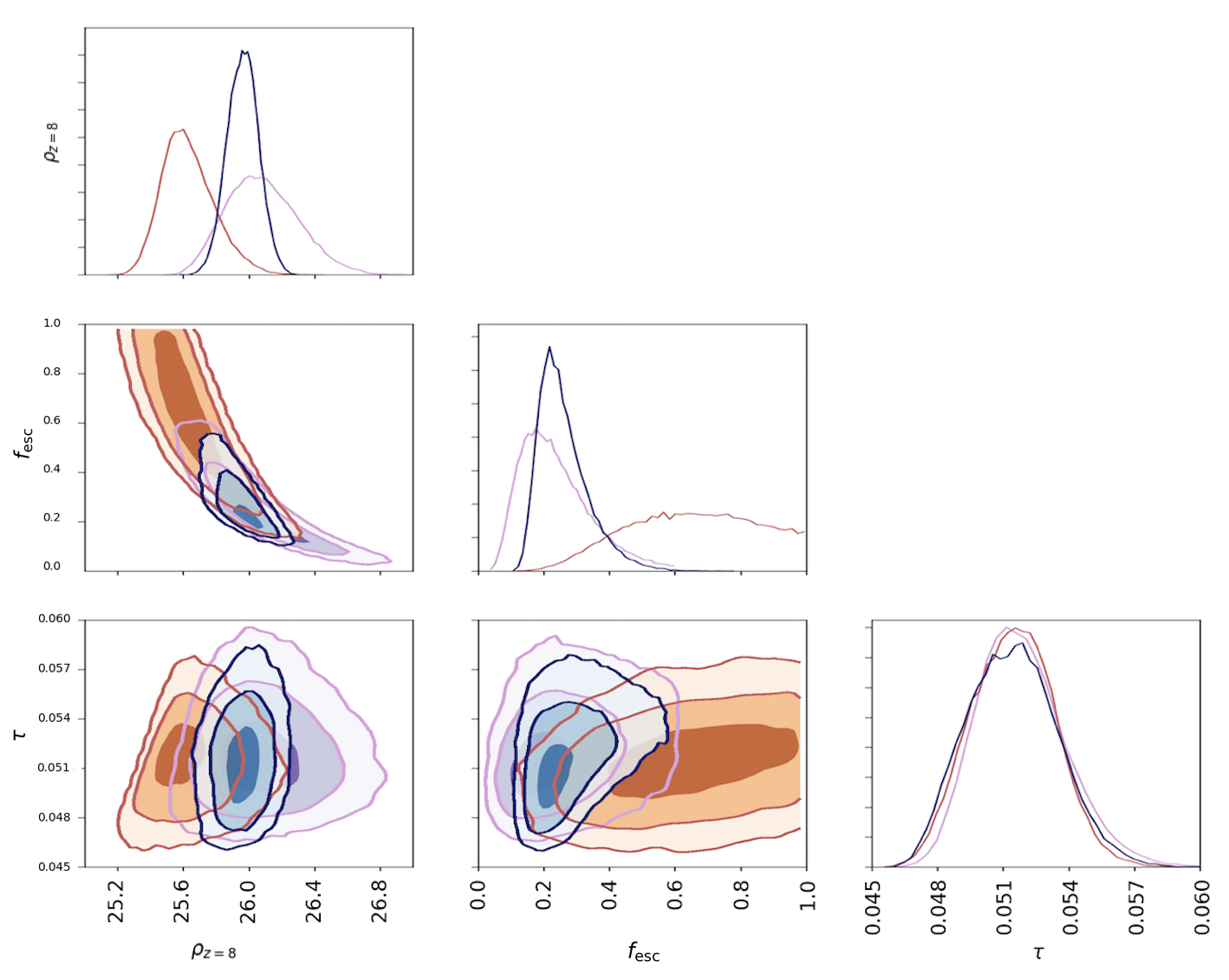}
\caption{Result of the MCMC analysis for the parametrisation described in Eq. \ref{eq:rho_Ishi} with $f_\mathrm{esc}$ added as a fitting parameter. Here, all three sets of observational data were used as constraints. The contours show the 1-, 2- and $3-\sigma$ confidence levels for log$(\rho_{z=8})$, $f_\mathrm{esc}$ and the derived parameter $\tau$. Colours correspond to the different values of the magnitude limit used in the analysis: purple for $M_\mathrm{lim}=-10$, blue for $M_\mathrm{lim}=-13$ and orange for $M_\mathrm{lim}=-17$. These results can be compared to reionisation histories displayed in Fig. \ref{fig:mag_comparison_QHII} and \ref{fig:mag_comparison_rhoSFR}.}
\label{fig:discussion_triangle_plot_fesc_Mlim}
\end{figure*}

However this comparison illustrates the systematic uncertainties on reionisation history due to the choice in the magnitude limit, but also in $f_\mathrm{esc}$ and $C_\ion{H}{II}$ values. We see these are much wider than the statistical uncertainties observed in Fig. \ref{fig:SFR_QHII_evolution} while still being reasonable. In particular, they mainly concern high redshifts. Indeed, we see in Fig. \ref{fig:mag_comparison_rhoSFR} that the $68\%$ confidence interval on star formation histories widens with redshift. However, few observations are available on this redshift range so we may expect that once data on earlier times is available, we will be able to improve constraints on the magnitude limit. In this perspective we can mention the work of \citet{mason_2017}, who derived a new constraint on reionisation history from simulations and models of the effects of IGM radiative transfer on Lyman-$\alpha$ emissions. They find an IGM ionised fraction at $z \sim 7$ of $x_\ion{H}{II} = 0.41 ^{+0.15} _{-0.11}$ in better agreement with our model for $M_{lim} = -17$ (see Fig. \ref{fig:mag_comparison_QHII}).

From a different point of view, \citet{price_2016} consider a varying
value of $M_\mathrm{lim}$ with redshift, and find that
$M_\mathrm{lim}$ varies in order to match the value of
$\tau_\mathrm{Planck}$ and to balance the increasing value of
$f_\mathrm{esc}$ with redshift allowed by their model. Here we find
that, overall, the model combines star formation history and ionised
fraction with difficulties when $M_\mathrm{lim}=-17$. Indeed,
Fig. \ref{fig:discussion_triangle_plot_fesc_Mlim} shows the
probability distribution functions of the parameters
log$(\rho_\mathrm{SFR})$ and $f_\mathrm{esc}$ and the corresponding
distribution of derived optical depths for the three choices of
magnitude limit. We see that for $M_\mathrm{lim}=-17$ the value of
$f_\mathrm{esc}$ is not well constrained and tends to be high. For
lower values of the escape fraction, the reionisation process needs to
start way earlier than in most of our results in order to have enough
radiation to fully ionise the IGM and to reach a sufficient value of
$\tau$. In fact, leaving the escape fraction as a free parameter
balances the uncertainty in the choice of $M_\mathrm{lim}$:
Fig. \ref{fig:mag_comparison_QHII} shows a narrower range of
uncertainties when we do not fix $f_\mathrm{esc}$, confirming the
correlation mentioned in \citet{price_2016}.

\subsection{Reionisation sources at $z>10$}
\label{subsec:discussion_sources}

Some doubts remain about the sources of reionisation: if \citet{robertson_2015} found that star-forming galaxies are sufficient to lead the process and to maintain the IGM ionised at $z \sim 7$ -- assuming $C_\ion{H}{II}=3$ and $f_\mathrm{esc}=0.2$, their analysis extrapolates luminosity functions between $z\simeq10$ and $z \simeq 30$, overlooking the possibility that other sources may have taken part in the early stages of reionisation process. Besides, they argue that low values of the Thomson optical depth reduce the need for a significant contribution of high-redshift galaxies and \citet{planck_2016} give much lower values than WMAP did \citep{hinshaw_2013}: $\tau_\mathrm{Planck}=0.058\pm0.012$ vs. $\tau_\mathrm{WMAP}=0.088\pm0.014$. Thus, now that we have investigated the possibility of this extrapolation, we chose to try the one of a constant SFR at $z\gtrsim10$. \\

We performed an MCMC maximum likelihood sampling of the 4-parameter model of $\rho_\mathrm{SFR}\,(z)$ in Eq. \ref{eq:rho_model} and add as a fifth parameter the the value of SFR density at $z>10.4$, our last data point corresponding to a redshift of $10.4$. We refer to it as $\rho_\mathrm{asympt}$ and chose to use all observations cited in Sect. \ref{subsec:observables_history} as constraints.
Final values of parameters $a$, $b$, $c,$ and $d$ are close to the ones from Sect. \ref{subsec:results_SFR}. We find that there is a strong correlation between $\rho_\mathrm{asympt}$ and $\tau$, because of the direct integration in Eq. \ref{eq:tau_def} and so expect higher values of the optical depth for high values of $\rho_\mathrm{asympt}$. Yet, $\tau$ values are limited by $Q_\ion{H}{II}$ data points and they have more impact on the global scenario. Indeed, models where $Q_\ion{H}{II}$ equals $30\%$ as soon as $z=10$ are allowed, whereas it is closer to $20\%$ at the same redshift when $\rho_\mathrm{SFR}$ is extrapolated. 
The correlation observed in our model parameters likelihood functions between $\rho_\mathrm{asympt}$ and $\tau$ had already been noticed by \citet{robertson_2015}, as a correlation between $\tau$ and the averaged value of $\rho_\mathrm{SFR}$ for $z>10$. A linear regression gives 
\begin{equation}
\label{eq:discussion_linear_tau_rho}
\mean{\rho_{\mathrm{SFR}}}_{z>10.4} = 0.51\, \tau - 0.026\ [\mathrm{M}_{\odot}\, \mathrm{yr}^{-1}\, \mathrm{Mpc}^{-3}],
\end{equation}
with a correlation coefficient $r=0.98$.

In this parametrisation, $\rho_\mathrm{asympt}$ can take very low values (down to $10^{-4}\ [\mathrm{M}_{\odot}\, \mathrm{yr}^{-1}\, \mathrm{Mpc}^{-3}]$) meaning that reionisation sources are almost completely absent at $z>10$. It also has an upper limit of $0.016\ [\mathrm{M}_{\odot}\, \mathrm{yr}^{-1}\, \mathrm{Mpc}^{-3}]$. This is close to the redshift-independent evolution of $\rho_\mathrm{SFR}$ ($\simeq10^{-1.5}\ [\mathrm{M}_{\odot}\, \mathrm{yr}^{-1}\, \mathrm{Mpc}^{-3}]$) considered by \citet{ishigaki_2015} for $z>3$ in order to reproduce $\tau_\mathrm{2014}=0.091\,^{+0.013}_{-0.014}$ \citep{planck_2013}, when usual decreasing models only gave them $\tau\simeq0.05$. We can compare Sect. \ref{subsec:results_SFR} results with this upper limit in Figure \ref{fig:SFR_rho_evolution}. Despite the wide range of possible values for $\rho_\mathrm{asympt}$, all results are consistent with our data and in particular, optical depths always remain in the $68\%$ confidence interval of $\tau_\mathrm{Planck}$. 

\subsection{How are $f_\mathrm{esc}$, $\dot{n}_\mathrm{ion}$ and $\rho_\mathrm{SFR}$ correlated?}
\label{subsec:discussion_correlations}

We expect a correlation between the amplitude $a$ of the star formation rate density parametrisation Eq. \ref{eq:rho_model} and the escape fraction. Indeed, $f_\mathrm{esc}$ takes no part in the estimation of $\rho_\mathrm{SFR}$ but they both take part in the calculation of $\dot{n}_\mathrm{ion}$ in Eq. \ref{eq:nion_mag} and then in the integration of $Q_\ion{H}{II}$ in Eq. \ref{eq:QHII_diff}. Thus, they must be constrained by the same data, so that the parameter $a$ can be a proxy for variations in the escape fraction value. To investigate this possible correlation, we plotted the distributions of $a \times f_\mathrm{esc}$  for various sets of constraints and in different models: with (\textbf{PAR}) and without (\textbf{CST}) the escape fraction as a fifth fit parameter and with all constraints.
We find that \textbf{CST} gives a lower value than \textbf{PAR} with a relative difference of $2.8\%$. This hints at a  correlation between $a$ and $f_\mathrm{esc}$ but more tests are needed to confirm or infirm this result. 

\begin{figure}
\centering
\includegraphics[width=0.5\textwidth]{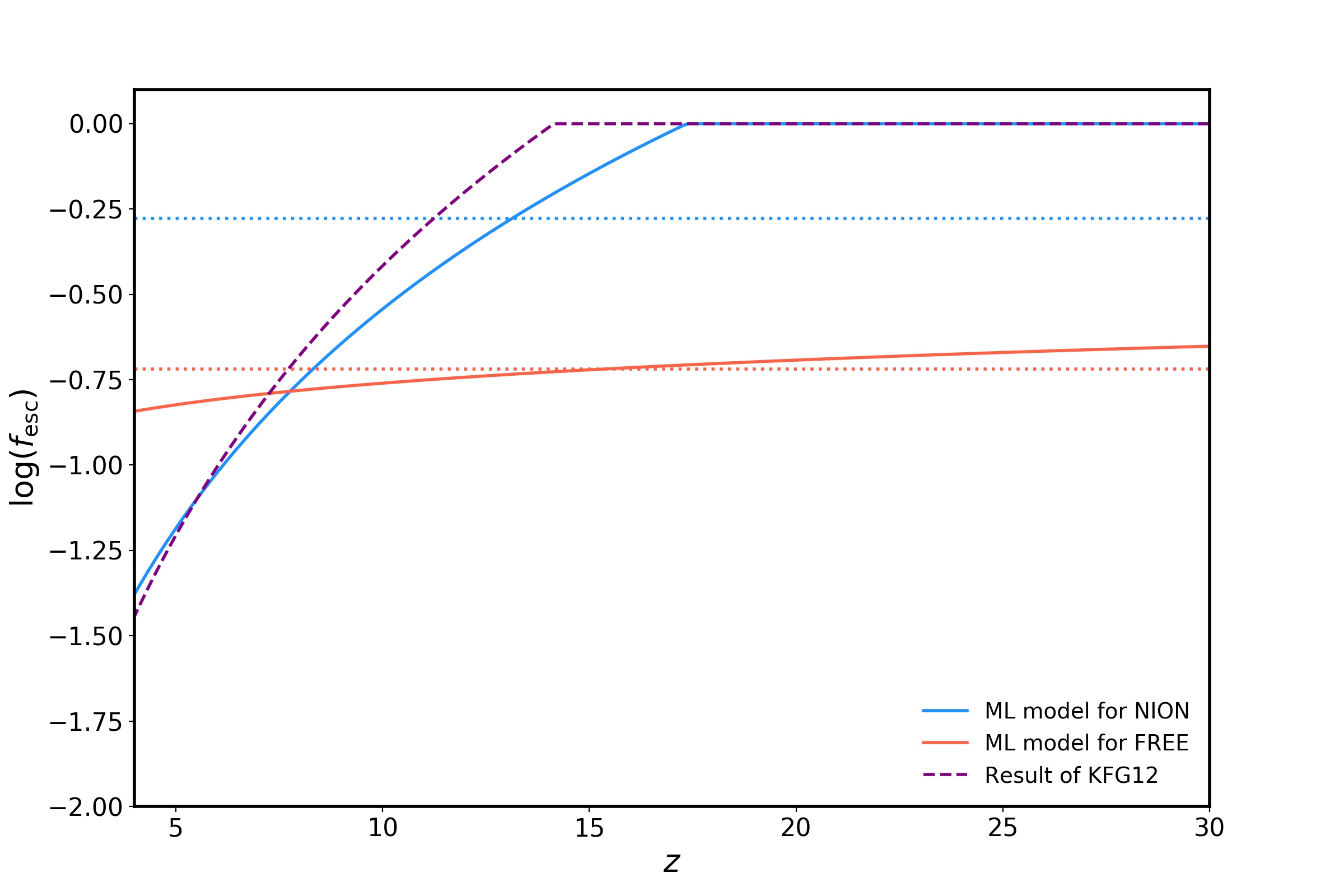}
\caption{Evolution of $f_\mathrm{esc}$ with redshift when $\dot{n}_\mathrm{ion}$ data points are used. ML models are shown for various set of constraints: blue when all constraints are used (\textbf{NION}); coral when $\dot{n}_\mathrm{ion}$ constraints are skipped (\textbf{FREE}). Horizontal dashed lines represent the mean value of $f_\mathrm{esc}$ over $4\leq z \leq30$ for the model of the corresponding colour. Inferences are compared to results of \citet[][KFG12, purple dashed line]{kuhlen_2012}.}
\label{fig:discussion_fesc_evolution}
\end{figure}

To further investigate the link between $f_\mathrm{esc}$, $\dot{n}_\mathrm{ion}$ and $\rho_\mathrm{SFR}$, we considered values of the reionisation rate  at various redshifts, used in \citet{kuhlen_2012} and \citet{robertson_2013}, and inferred from measurements and calculations of \citet{faucher-giguere_2008,prochaska_2009,songaila_2010}. We call \textbf{NION} the run using these new constraints -- in addition to the others -- and \textbf{FREE} the one skipping them, corresponding to \textbf{ALL} from Sect. \ref{subsec:results_fesc}.

We compare in Table \ref{table:discussion_nion_comparison} values of
the reionisation rate at various redshifts for \textbf{NION},
\textbf{FREE} and \citet{kuhlen_2012}. \textbf{NION} gives results
close to data points, increasing with $z$, whereas \textbf{FREE}
values are significantly higher and decrease with redshift. This
difference in the evolutions of $\dot{n}_\mathrm{ion}\,(z)$ is
directly linked to $f_\mathrm{esc}\,(z)$. We see in Fig. \ref{fig:discussion_fesc_evolution} that
  $f_\mathrm{esc,FREE}$ is almost constant with redshift and therefore $\dot{n}_\mathrm{ion}\,(z)$ decreases on this redshift range: because $f_\mathrm{esc}$ values remains quite high, there is no need for many ionising sources at high redshift.  On the contrary, when the constraints on $\dot{n}_\mathrm{ion}$ are included in the fit, the reionisation rate takes overall lower values (see Table \ref{table:discussion_nion_comparison}) so that $f_\mathrm{esc,NION}$ has to take higher values at high redshift (saturating to 1 for $z \geq 15$) to compensate for the lack of ionising sources. However, this is still hardly sufficient and we find that for \textbf{NION}, the reionisation process needs to start as early as at $z =8$ to fully ionise the IGM, with $Q_\ion{H}{II}=1$ being reached later than others models, around $z=5.5$. This behaviour leads to a high value of $\tau = 0.082$, at the edge of the
  3-$\sigma$ confidence interval of $\tau_\mathrm{Planck}$ and
  therefore hardly compatible with observational results \citep{planck_2016}. Removing the
  constraints on the filling factor, $f_\mathrm{esc}$ remains low on
  the whole redshift range ($<20$). We then get values of the optical
  depth in agreement with Planck ($0.058\pm0.011$) but reionisation
  does not end before $z\sim4$. Thus, the estimations on the
  reionisation rate from
  \citet{faucher-giguere_2008,prochaska_2009,songaila_2010} are
  compatible with one observable at a time: either the ionisation
  level -- leading to a higher value of $\tau$ --, or the Thomson
  optical depth -- so that reionisation ends around $z \sim 4$ -- ,
  but cannot match all observations in a coherent way.
  
\begin{table}
\caption{Comparison between our results and data points on the cosmic reionisation rate from \citet[][KFG12]{kuhlen_2012}.}
\label{table:discussion_nion_comparison}
\centering
\begin{tabular}{ccccc}
\hline \hline
$z$ & \multicolumn{3}{c}{$\dot{n}_\mathrm{ion}\ [\,10^{50}\ \mathrm{s}^{-1}\, \mathrm{Mpc}^{-3}\,]$}\\
 \hline
 & KFG12 & NION & FREE \\ 
\hline
$4.0$ & $3.2_{-1.9}^{+2.2}$ & $3.9 \pm 0.7$ & $13.7 ^{+3.9}_{-5.1}$ \\ 
$4.2$ & $3.5_{-2.2}^{+2.9}$ & $4.0  \pm 0.7$ & $12.6 ^{+3.5}_{-4.5}$ \\ 
$5.0$ & $4.3 \pm 2.6$ & $4.1 \pm 0.5$ & $9.3 ^{+2.3}_{-2.6}$ \\ 
 \hline
\end{tabular} 
\end{table}

\begin{table*}[h]
\caption{ML parameters from the fit on $\rho_\mathrm{SFR}$ with various parameters and constraints.} \label{table:results_rho_parameters}      
\centering          
\begin{tabular}{c c c c c c c c c c } 
\hline\hline       
Ref. & \multicolumn{3}{c}{Constraints} & \multicolumn{4}{c}{$\rho_\mathrm{SFR}$ parameters} &  \multicolumn{2}{c}{Other parameters} \\ 
\hline     
         & $\rho_\mathrm{SFR}$ & $Q_\ion{H}{II}$ & $\tau_\mathrm{Planck}$ & $a$ & $b$ & $c$ & $d$ & $f_\mathrm{esc}$ & $C_\ion{H}{II}$ \\
\hline
        ALL & \cmark & \cmark & \cmark & $0.0146\pm0.0011$ & $3.17\pm0.20$ & $2.65\pm0.14$ & $5.64\pm0.14$ & -- & -- \\ 
         & \cmark & \xmark & \xmark & $0.0145\pm0.0011$ & $3.20\pm0.22$ & $2.63\pm0.15$ & $5.68\pm0.19$ & -- & -- \\  
        NORHO & \xmark & \cmark & \cmark & $0.0129\pm0.343$ & $0.458\pm0.970$ & $5.69\pm1.65$ & $7.14\pm1.90$ & -- & -- \\
         & \cmark & \cmark & \xmark & $0.0147\pm0.0011$ & $3.17\pm0.21$ & $2.66\pm0.14$ & $5.63\pm0.14$ & -- & -- \\  
        NOQ & \cmark & \xmark & \cmark & $0.0145\pm0.0011$ & $3.22\pm0.22$ & $2.61\pm0.15$ & $5.66\pm0.19$ & -- & -- \\  
\hline
        ALL & \cmark & \cmark & \cmark & $0.0147\pm0.0011$ & $3.14\pm0.21$ & $2.69\pm0.15$ & $5.74\pm0.19$ & $0.193\pm0.037$ & -- \\  
        NOQ & \cmark & \xmark & \cmark & $0.0146\pm0.0011$ & $3.18\pm0.21$ & $2.65\pm0.15$ & $5.70\pm0.19$ & $0.213\pm0.079$ & -- \\  
\hline         
        ALL & \cmark & \cmark & \cmark & $0.0146\pm0.0011$ & $3.18\pm0.21$ & $2.65\pm0.15$ & $5.67\pm0.19$ & -- & $4.56\pm1.85$ \\  
        NOQ & \cmark & \xmark & \cmark & $0.0145\pm0.0012$ & $3.20\pm0.22$ & $2.63\pm0.15$ & $5.69\pm0.19$ & -- & $5.10\pm2.74$ \\  
        \hline
        ALL$^*$ & \cmark & \cmark & \cmark & $0.0147\pm0.0011$ & $3.14\pm0.21$ & $2.69\pm0.15$ & $5.75\pm0.19$ & $0.20 \pm 0.05$ & $3.50\pm1.10$ \\
        \hline

\end{tabular}
\tablefoot{$^*$: Prior on $f_\mathrm{esc}$ and $C_\ion{H}{II}$ are different for comparison with \citet{price_2016} -- see text for details.} 
\end{table*}

\begin{table*}[h]
\caption{ML parameters for the fits on $f_\mathrm{esc}(z)$ and $C_\ion{H}{II}(z)$ in, respectively, Sect. \ref{subsec:results_fesc} and \ref{subsec:results_ch2}.}             
\label{table:results_fesc_ch2_parameters}      
\centering          
\begin{tabular}{c c c c c c c c } 
\hline\hline       
Model & Reference & $Q_\ion{H}{II}$ & $\tau_\mathrm{Planck}$ & \multicolumn{4}{c}{Model parameters} \\ 
\hline     
         \multirow{4}*{$f_\mathrm{esc}(z)$} & \multirow{4}*{KFG12} &  &  & \multicolumn{2}{c}{$\alpha$} & \multicolumn{2}{c}{$\beta$} \\
         & & \cmark & \cmark & \multicolumn{2}{c}{$0.14\pm0.02$}& \multicolumn{2}{c}{$0\pm0.29$} \\ 
         & & \cmark & \xmark & \multicolumn{2}{c}{$0.15\pm0.02$}& \multicolumn{2}{c}{$0\pm0.30$} \\
         & & \xmark & \cmark & \multicolumn{2}{c}{$0.11\pm0.09$}& \multicolumn{2}{c}{$0\pm0.78$} \\ 
\hline  
         \multirow{5}*{$C_\ion{H}{II}(z)$} & &  &  & $\alpha$ & $a$ & $b$ & c \\   & \multirow{2}*{HM12} & \cmark & \cmark & $0.74 \pm 0.29$ & $5.74\pm1.07$ & $-1.21 \pm0.58$ & -- \\
         & & \cmark & \xmark & $0.79 \pm 0.29$ & $5.56\pm1.09$ & $-1.30\pm0.69$ & -- \\             
         & \multirow{2}*{I07} & \cmark & \cmark & -- & $7.29\pm1.63$ & $-0.042\pm0.030$ & $0\pm 2.4\times 10^{-4}$ \\
         & & \cmark & \xmark & -- & $7.11\pm1.17$ & $-0.046\pm0.058$ & $0\pm 6.3\times10^{-4}$ \\
         \hline
\end{tabular}
\tablebib{KFG12: \citet{kuhlen_2012}; HM12.1 $\&$ HM12.2: \citet{haardt_2011}; I07: \citet{iliev_2007}.
}
\end{table*}

\section{Conclusions}
\label{sec:conclusions}

We used the latest observational data available on reionisation history, i.e. cosmic star formation density, ionised fraction of the IGM and Thomson optical depth derived from Planck observations to find that they are all compatible with a simple and credible scenario where reionisation begins around $z=15$ and ends by $z=6$. Among all data, star formation history seems to be the most constraining for the EoR.

An investigation of various parametrisations of the escape fraction of ionising photons has lead us to conclude that it is very well constrained by observations: when considered constant with redshift, values allowed by the fit range from $20\%$ to $28\%$; when considered redshift-dependent, from $f_\mathrm{esc}\,(z=4)\simeq17\%$ to $f_\mathrm{esc}\,(z=30)\simeq26\%$ following a low increase with $z$. The fiducial constant value of $20\%$ often used in papers seems then to be perfectly consistent with our data. {However, one must keep in mind that these results strongly depend on the hypothesis we make about the magnitude limit as a lower value of $M_\mathrm{lim}$ will require higher values of $f_\mathrm{esc}$ and vice versa. While the constraints on $\tau$ are unaffected by the assumption on $M_\mathrm{lim}$, the confidence range on $f_\mathrm{esc}$  is enlarged for $M_\mathrm{lim}=10$.  Furthermore, our different sets of observations seem to be in tension with each other for $M_\mathrm{lim}=-17$ or for values of $f_\mathrm{esc} \lesssim 10\%$.}

On the contrary, the clumping factor of ionised hydrogen in the IGM can take a wide range of different values without impacting the reionisation observables significantly. For instance, when take $C_\ion{H}{II}$ as a redshift-independent parameter, its relative standard deviation is $41\%$ whereas it is at most $7.6\%$ for $Q_\ion{H}{II}\,(z)$\footnote{Reached at $z=6.2$.}. The result is the same when we consider that $C_\ion{H}{II}$ depends on redshift: a great variety of possible evolutions gives the same scenario in terms of ionisation level. There is no greater impact on Thomson optical depth values, which vary of a maximum of a few percent compared to $\left\langle \tau \right\rangle _{C_\ion{H}{II}=3}$ and always remains in the 1-$\sigma$ confidence interval of $\tau_\mathrm{Planck}$. Observational constraints are thus extremely robust to variations of the clumping factor. We nevertheless find a correlation between the averaged value of $C_\ion{H}{II}$ for $z \in [6.8,30]$ and $\tau$: the linear fit
\begin{equation}
\label{eq:cl_lin_ch2_tau}
\mean{C_\ion{H}{II}}_{z>6.8} = -350\, \tau +24.4
\end{equation}
provides a good description of their connection\footnote{Here, the model from Eq. \ref{eq:ch2_model_hm} was considered.}. This supports the use of a redshift-independent clumping factor to study the EoR. A possible choice, consistent with observations, would then be $C_\ion{H}{II}=3$, the fiducial value often used in papers, because it lies in the range of the ML $C_\ion{H}{II}$ values found in Sect. \ref{subsec:results_ch2}.

Last, a quick study on the possible reionisation sources at $z\gtrsim10$ showed that there is no need for exotic sources such as early quasars \citep{madau_2015} or for an artificial increase in star formation density at high redshift \citep{ishigaki_2015}. When their luminosity functions are extrapolated, a hypothesis still recently strongly supported by \citet{livermore_2016}, star-forming galaxies provide enough photons to have a fully ionised IGM at $z=6$.

\begin{acknowledgements}
The authors thank B.E. Robertson for kindly providing us with his compilation of star formation rate densities. They thank the referee for useful comments. This research made use of Astropy, a community-developed core Python package for Astronomy \citep{astropy,astropy2}; matplotlib, a Python library for publication quality graphics \citep{hunter_2007} and emcee, an implementation of the affine invariant MCMC ensemble sampler \citep{emcee}. This work was partly supported by Programme National de Cosmologie et Galaxies (PNCG). AG acknowledges financial support from the European Research Council under ERC grant number 638743-FIRSTDAWN as well as from an STFC PhD studentship. 
\end{acknowledgements}

\appendix

\section{MCMC multidimensional plots}

We show in this appendix the additional triangle plots of the runs ALL corresponding to the studies with $f_{esc}$ as additional free parameter (see Section \ref{subsec:results_SFR}), with $C_{HII}$ as additional free parameter (see Section \ref{subsec:results_ch2}), and finally with both free (see Section \ref{subsec:fesc_Ch2}).

\begin{figure}
\resizebox{\hsize}{!}{\includegraphics{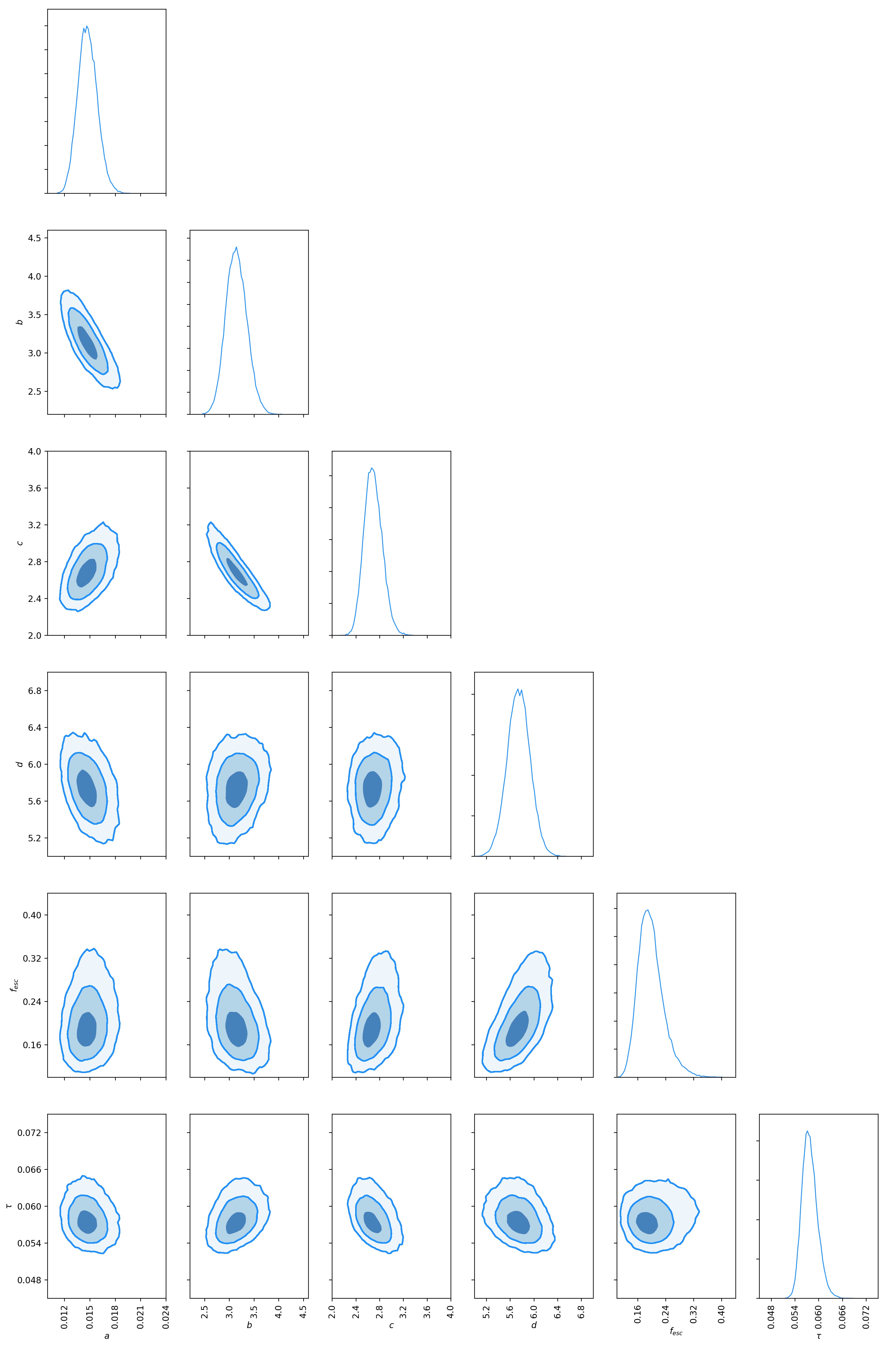}}
\caption{Results of the MCMC analysis for the ALL case when $f_{esc}$ is added as a free parameter. The contours show the 1, 2, and 3  $\sigma$confidence levels for a, b, c, d,  $f_{esc,}$ and the derived parameter $\tau$.}
\label{fig:Rfesc_triangle}
\end{figure}

\begin{figure}
\resizebox{\hsize}{!}{\includegraphics{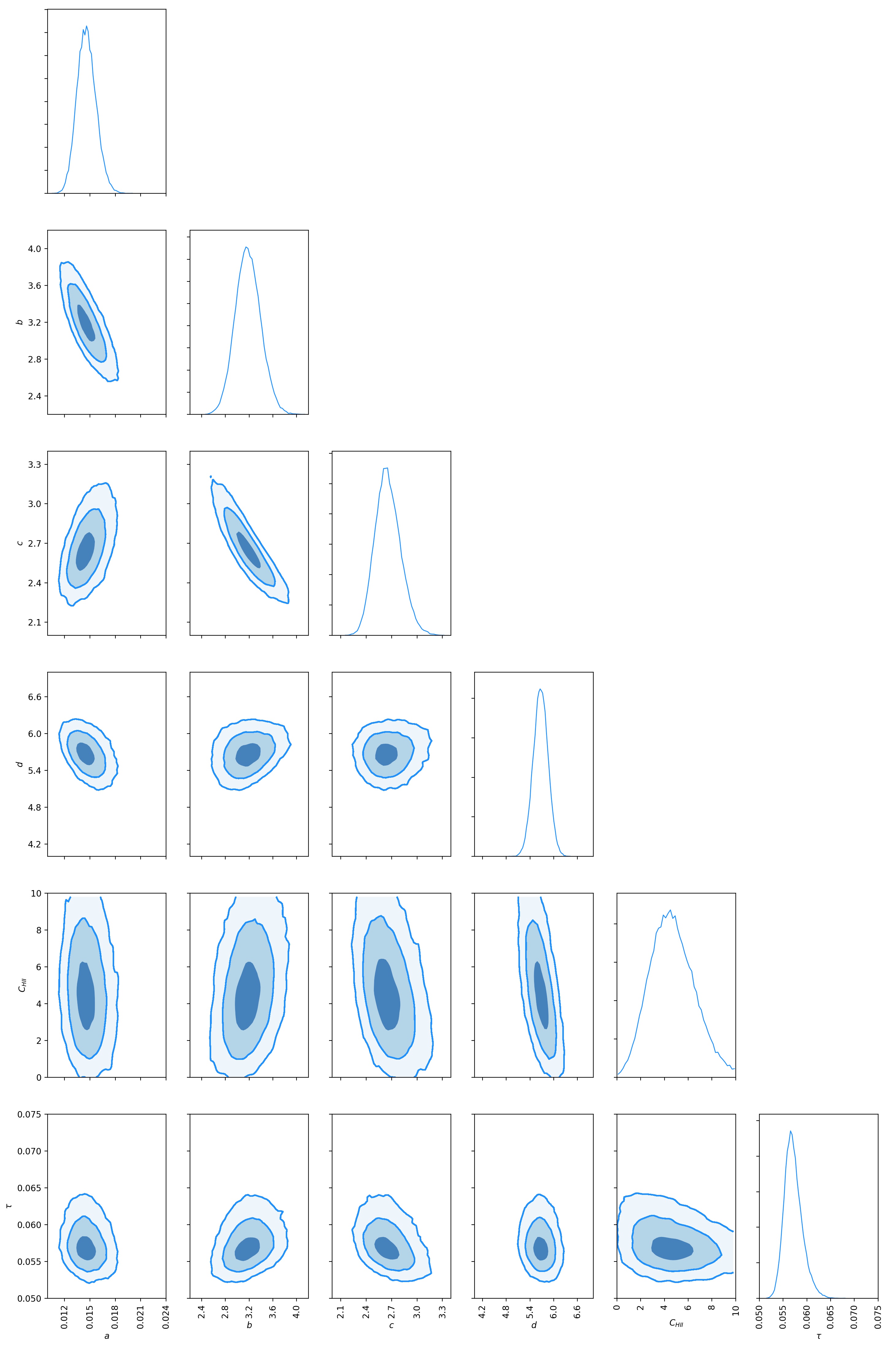}}
\caption{Results of the MCMC analysis for the ALL case when $C_{HII}$ is added as a free parameter. The contours show the 1, 2, and 3 $\sigma$ confidence levels for a, b, c, d, $C_{HII,}$ and the derived parameter $\tau$.}
\label{fig:RCh2_triangle}
\end{figure}

\begin{figure}
\resizebox{\hsize}{!}{\includegraphics{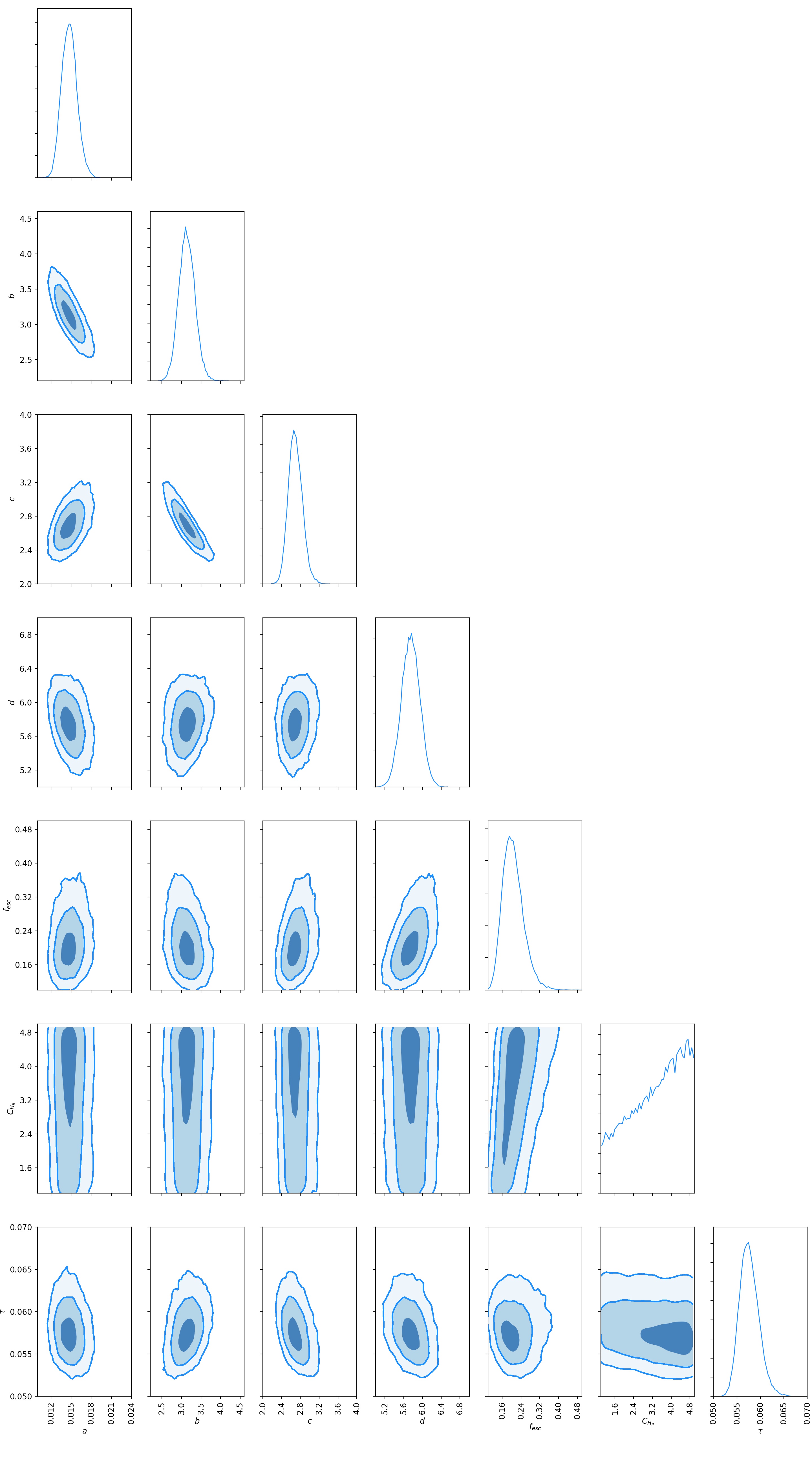}}
\caption{Results of the MCMC analysis for the ALL case when both $f_{esc}$ and $C_{HII}$  are added as a free parameter. The contours show the 1, 2, and 3 $\sigma$ confidence levels for a, b, c, d,  $f_{esc}$, $C_\ion{H}{II,}$ and the derived parameter $\tau$.}
\label{fig:RfescCh2_triangle}
\end{figure}

\bibliographystyle{aa} 
\bibliography{biblio} 

\end{document}